\documentclass[%
 reprint,
 amsmath,amssymb,
 aps,
pra,
]{revtex4-1}

\usepackage{graphicx}
\graphicspath{ {Figures/} }

\usepackage{dcolumn}
\usepackage{bm}
\usepackage{bbold}
\usepackage[export]{adjustbox}

\usepackage{mathtools} 
\DeclarePairedDelimiter\bra{\langle}{\rvert}
\DeclarePairedDelimiter\ket{\lvert}{\rangle}
\DeclarePairedDelimiterX\braket[2]{\langle}{\rangle}{#1 \delimsize\vert #2}

\usepackage[utf8]{inputenc}
\usepackage[T1]{fontenc}

\usepackage{mathrsfs}
\usepackage{dsfont}

\usepackage{amssymb}
\usepackage{amsmath}
\usepackage{amsfonts}
\usepackage{amsthm}

\usepackage{hyperref}
\hypersetup{pdfpagemode=UseNone}

\newcounter{rem}
\newcommand{\mc}[1]{\mathcal{#1}}

\def\>{\rangle}
\def\<{\langle}
\newcommand{\proj}[1]{| #1 \rangle\! \langle #1 |}
\renewcommand{\rho}{\varrho}

\def\tr{{\rm Tr}}

\def\ii{{\rm i}}

\def\textbf#1{{\bf #1}}

\def\beq{\begin{equation}}
\def\eeq{\end{equation}}
\def\beqa{\begin{eqnarray}}
\def\eeqa{\end{eqnarray}}
\usepackage{xcolor}

\begin{document}

\title{Spin-entanglement wave in a coarse-grained optical lattice}

\author{Pedro Silva Correia}
\email{pedrosc8@cbpf.br}
\affiliation{Centro Brasileiro de Pesquisas F\'isicas, \\ Rua Dr.~Xavier Sigaud, 150, Rio de Janeiro, RJ, Brasil}

\author{Fernando de Melo}
\email{fmelo@cbpf.br}
\affiliation{Centro Brasileiro de Pesquisas F\'isicas, \\ Rua Dr.~Xavier Sigaud, 150, Rio de Janeiro, RJ, Brasil}

\date{\today}

\begin{abstract}
In the present work we explore a suitable coarse graining channel as a tool to describe the effective entanglement spreading in a coarse-grained spin-chain with different degrees of resolution. Comparing it with the experimental realizations performed with ultracold atoms, our results suggest that even if we are not able to fully resolve the system, entanglement can still be detected for some coarse graining levels. Furthermore, we show that it is possible to have some information about the ``microscopic'' entanglement, even if we have access only to the system's coarse-grained description. We show that the amount of entanglement decays exponentially with the lack of system resolution. The lack of experimental resolution might thus lead to an effective classical description.
\end{abstract}


\maketitle

\section{Introduction}

The 2012 Nobel Prize was awarded to Serge Haroche and David J.~Wineland for ``ground-breaking experimental methods that enabled measuring and manipulation of individual atoms''~\cite{haroche,wineland}. Present day quantum technologies have evolved as to allow for measurement and control of  well-isolated quantum systems with an increasing number of qubits~\cite{google,ibm,dwave}. However, such an advance comes with an crucial difficulty. As the size of the system increases, the experimental control and characterization of many-body quantum systems spend expensive resources, making them impracticable. For instance, brute force quantum state tomography requires a challenging individual particle access, with both experimental procedure and data post-processing inherently time-consuming for large systems~\cite{haffner2005scalable}. Nevertheless, quantum systems may not need to be completely resolved in order to display some genuine quantum signature. Imprecise measuring devices may then still be useful in characterizing some aspects of quantum many-body systems. A question then arises: How imprecise can a measurement be  while still allowing for some quantum feature to be detected? Here we introduce an approach that aims at describing effective quantum systems, i.e., a coarse graining description of the quantum state in which the system's degrees of freedom that cannot be resolved by  the measurement apparatus are directly discarded. With such an effective description we can then determine if genuine quantum features can still be observed. 

As a genuine quantum signature, here we focus on the multi-particle quantum entanglement. Quantum entanglement is an especial kind of correlation present in quantum many-body systems that plays an essential role in the quantitative description of strong correlated regimes. Entanglement is fragile in its nature and it is absent in the classical world~\cite{aolita}. Usually this fragility is taken into account by the decoherence theory, which describes the loss of information about the quantum system due to its interaction with the environment~\cite{aolita}.  There are scenarios, nonetheless, in which the fragility of entanglement is intrinsic to the level of description that we have access to -- be it in order to avoid the colossal complexity of a complete specification, or due to the non-perfect resolution of the measurement apparatus used to characterize the system. In this sense, we will analyze up to what level of description, as modeled here by a blurred detector, entanglement can still be measured. Thus, in addition to understand the impact of a coarse-grained description on entanglement detection, we hope that this approach may shed some light on the classical-to-quantum transition, bringing to the discussion other mechanisms that perturb quantum resources.

Despite of the generality of our approach, to make its application more concrete, here we look at a recent experimental realization exploring many-body entanglement: spin systems realized with ultracold atoms in optical lattices~\cite{gross2017quantum}.  Equipped with a high-resolution quantum gas microscope~\cite{sherson,gross2017quantum}, it was possible to measure spin-entanglement waves in a few qubits Bose-Hubbard chain~\cite{fukuhara,fukuhara2}. Such a measurement process is based on fluorescence technique: the spin-chain is illuminated with a laser in way that the microscope detects scattered light by atoms in excited states. Thinking in this experimental situation together with its theoretical description~\cite{spindynamics1,spindynamics2}, we construct a coarse graining map that models a microscope with an insufficient resolution~\cite{duarte,PedroM2017}.  Although our construction is inspired by the fluorescence imaging -- being thus directly applicable to many other experimental platforms \cite{atomdet} -- other coarse graining maps can be designed as to model other experimental detection setups.

Our coarse graining approach is somewhat inspired on Kadanoff's blocking procedure that initiated the renormalization theory~\cite{kadanoff,wilson}, and it is also slightly related to the more recent entanglement renormalization procedure proposed by Vidal~\cite{vidal2007entanglement}. Like in these two methods, here the effective system also has a smaller  description. Differently, however, we do not try to optimize the effective description as to maintain intact, for instance, the value of local observables. Rather, here we take a more experimentally motivated venue and search for  descriptions that can be actually measured in a given experimental realization. Our coarse graining map thus models an incoherent loss of information: a pure microscopic state is seen as an  effective  macroscopic mixed state. In this sense, the use of a coarse graining map (a general completely positive map that reduces the system dimension) is a generalization of decoherence theory (related to the partial trace map),  allowing us, in the present application, to describe the entanglement behavior taking into account different ranges of resolution. By applying the coarse graining map proposed in~\cite{duarte,PedroM2017}, we will be able to observe the decay of entanglement dynamics in different degrees of the lattice sharpness and see to what extent its detection is still achievable. Recently, in Ref.~\cite{Ibrahim}, a similar approach is used in an abstract way in order to derive entanglement measures for high-dimensional bipartite systems.

This paper is organized as follows: In Sec. \ref{sec:entspinchain}, we present a brief description of the entanglement spreading during single spin-impurity dynamics in one-dimensional Bose-Hubbard system. After that, in Sec. \ref{sec:cg} we define a general coarse graining operation as a completely positive trace-preserving (CPTP) map that reduces the system's dimension. Then, inspired by fluorescence imaging of atoms in an optical lattice, we present a coarse graining map that describes a spin detection with insufficient resolution. In Sec. \ref{sec:cgspinchain} we use this coarse graining map to describe the entanglement due to spin-impurity dynamics in a coarse-grained Bose-Hubbard spin-chain. We explore how entanglement behaves when we take into account different ranges of resolutions of the spin-chain. Finally in Sec. \ref{sec:conclusions} we summarize our results and discuss some implications of this coarse graining approach.

\section{Entanglement in a Bose-Hubbard spin-chain}
\label{sec:entspinchain}

In this section we  describe the entanglement generation and spreading during spin-impurity dynamics in a 1D spin-$1/2$ $XX$-chain \cite{spindynamics1,spindynamics2}. In the experimental work \cite{fukuhara,fukuhara2}, it was produced a ferromagnetic Heisenberg spin-chain with ultracold bosonic atoms in an optical lattice. In these experiments, individuals atoms are trapped in each potential minimum of a periodic potential associated with a stationary wave created by counter-propagating laser beams \cite{bloch,bloch2}. Deep in the Mott-insulator regime with unity filling, two hyperfine levels of each atom act as a spin-1/2 (qubit), and neighboring spins interact with each other via isotropic spin-$1/2$ Heisenberg Hamiltonian \cite{heisenbergxx}: $\hat{H}=-J_{ex}\sum_j\hat{\textbf{\sigma}}_j\cdot\hat{\textbf{\sigma}}_{j+1}$. In this Hamiltonian,  $\hat{\textbf{\sigma}}_j=(\hat{\sigma}_j^x,\hat{\sigma}_j^y,\hat{\sigma}_j^z)$ denotes the spin-$1/2$ vector of Pauli matrices at site $j$, and $J_{ex}$ is the exchange coupling which is constant for homogeneous chains (see the supplementary information of \cite{fukuhara2}). In the case of a single spin-impurity in a 1D lattice (single excitation subspace), the Hamiltonian can be written in a simplified form:
\begin{equation}
\hat{H}=-J_{ex}\sum_j(\hat{\sigma}_j^+\hat{\sigma}_{j+1}^- + \hat{\sigma}_j^-\hat{\sigma}_{j+1}^+),
\label{eq:HXX}
\end{equation}
where $\hat{\sigma}_j^\pm=(\hat{\sigma}_j^x{\pm}\ii\hat{\sigma}_j^y)/2$ are the spin-$1/2$ raising (lowering) operators. The term $J_{ex}\sum_j\hat{\sigma}_j^z\hat{\sigma}_{j+1}^z$ was dropped, since it gives rise only to an energy offset within the single excitation subspace~\cite{fukuhara2}.

A infinite spin-up chain with a single spin-down (spin-impurity) on site $j$ can be represented by the state
\begin{equation}
\ket{j}\equiv\ket{\cdots, 0_{j-1}, 1_j, 0_{j+1}, \cdots},
\label{eq:chain}
\end{equation}
where $\ket{1}$ and $\ket{0}$ refers to spin down and spin up states respectively, in the $z$-basis. As initial state it is considered a single spin-down at the ``center'' of the  chain ($j=0$). The spin-impurity spreading is given by the time evolution generated by the Hamiltonian in equation (\ref{eq:HXX}), and it can be described by 
\begin{equation}
\ket{\psi(t)}=\sum_j\phi_j(t)\ket{j},
\label{eq:impurityspread}
\end{equation}
where $\phi_j(t)=\ii^jJ_j(J_{ex}t/\hbar)$, with  $J_j(x)$ the Bessel function of the first kind~\cite{bessel}. For simplicity, from now on we will consider the time evolution in dimensionless time unit $J_{ex}t/\hbar$.

\subsection{Concurrence between two sites}

The next step is to quantify the entanglement between spins in two different sites $A$ and $B$ in the chain. 
The reduced density operator related to a pair of different arbitrary sites $A$ and $B$ is given by:
\begin{equation}
\psi_{AB}(t)=\tr_{\overline{AB}}[\ket{\psi(t)}\!\bra{\psi(t)}],
\end{equation}
where $\overline{AB}$ means the complementary sites to $AB$.
Using (\ref{eq:impurityspread}) and the basis states $\ket{00}$, $\ket{01}$, $\ket{10}$ and $\ket{11}$ for sites $A$ and $B$, we get:
\begin{equation}
\psi_{AB}=\begin{pmatrix}
1-|\phi_A|^2-|\phi_B|^2 &            0        &          0          &           0             \\
0 &       |\phi_B|^2    & \phi_A\phi_B^{\ast} &           0             \\
0 & \phi_A^{\ast}\phi_B &     |\phi_A|^2      &           0             \\
0 &            0        &          0          & 0
\end{pmatrix}.
\label{eq:rhoAB}
\end{equation}
Explicit time dependence is suppressed whenever obvious, to avoid cluttered notation.

\begin{figure}[h!]
	\begin{tabular}{rc}
		(a) &\hspace*{0.8cm} \includegraphics[width=5.5cm ,valign=t]{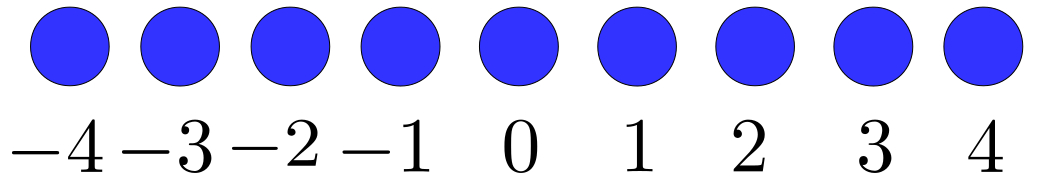} \\
		 (b)&\includegraphics[width=0.8\linewidth ,valign=t]{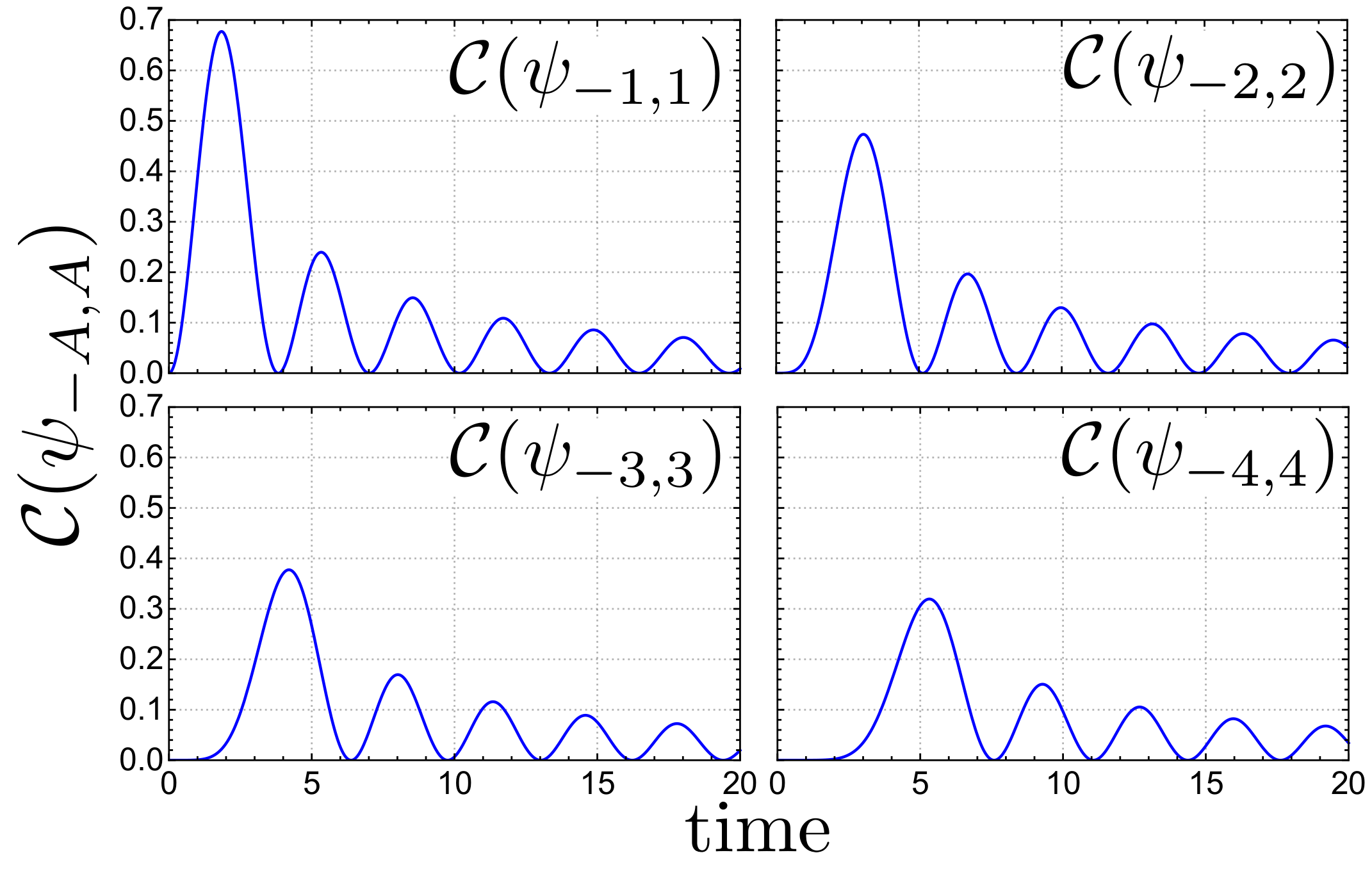} \\
		 (c) & \includegraphics[width=0.8\linewidth,valign=t]{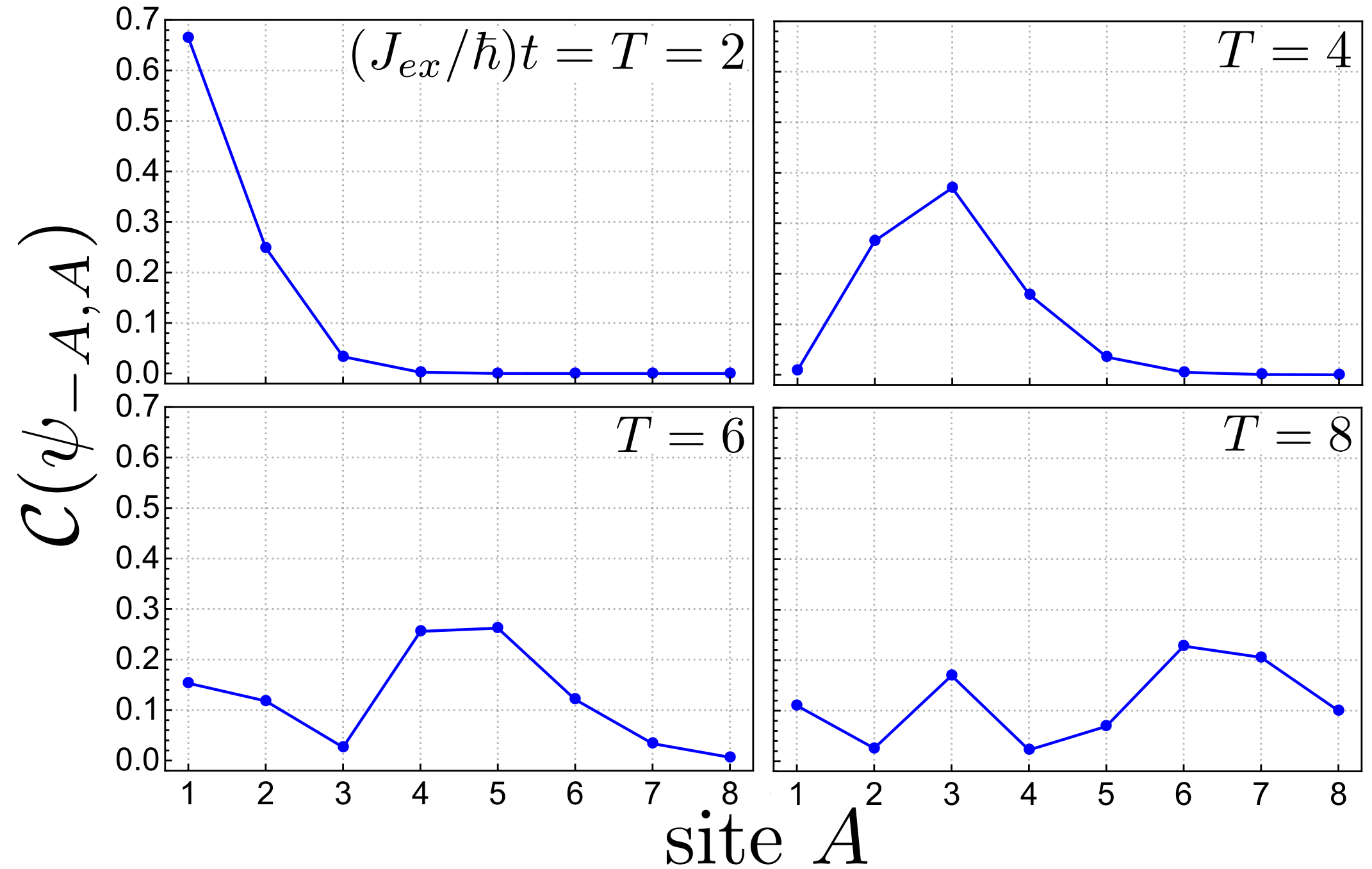}
	\end{tabular}
    \caption{\footnotesize \textbf{Entanglement dynamics of a single impurity spin-chain.} (a)  Schematic representation of the spin-chain with its labels (the time evolution is given in dimensionless unit $J_{ex}t/\hbar$). (b) Time evolution of concurrence $\mathcal{C}(\psi_{A,-A})$ between spins of sites $A$ and $-A$ for different choices of $A$ (c) Spin entanglement wave with single site resolution.  Time is given in dimensionless unit $J_{ex}t/\hbar$}
    \label{fig:concevolution}
\end{figure}

The matrix representation (\ref{eq:rhoAB}) of $\psi_{AB}$ can be identified as a $X$-matrix,
whose concurrence can be easily calculated~\cite{concX} to give:
\begin{equation}
\mathcal{C}(\psi_{AB})=2|\phi_A\phi_B^{\ast}|.
\label{eq:concAB}
\end{equation}
Figure~\ref{fig:concevolution} illustrates the expected entanglement dynamics between symmetric sites with respect to the spin at position $j=0$. Such an entanglement wave -- see Fig.~\ref{fig:concevolution}(b) -- was observed experimentally showing a reasonable agreement with the theoretical prediction \cite{fukuhara}.

\section{Coarse-grained Description of a Spin Quantum System}
\label{sec:cg}

Following a scheme proposed in \cite{mazza} and supported by an experimental setup with single-site-resolved fluorescence imaging \cite{sherson,gross2017quantum}, researchers have detected a lower bound for the concurrence of two sites. Their results are consistent with those presented in Fig. \ref{fig:concevolution}. The single-site-resolved fluorescence imaging is roughly described as follows: each atom of the chain is illuminated with a laser in way that if light is scattered, the atom was in state $\ket{0}$; whereas if no light is scattered, the atom was in state $\ket{1}$. To detect the light coming from each single atom, a high resolution quantum gas microscope is required (single-site-resolution scheme \cite{sherson}). In the case of a truly many-body system such equipment may not be able to resolve a single atom, what may become an obstacle in order to observe the entanglement dynamics. With that in mind, we propose a coarse-grained scheme to simulate a detector with insufficient resolution and we show how entanglement behaves in that situation.

The essential idea of coarse graining models is to give an effective description of a complex system focusing on a small number of degrees of freedom, i.e., ignoring the system's finer details. In this way, coarse graining models can be used as tool to explain the emergence of large scale structures starting from an underlying microscopic theory.Within  quantum mechanics, a coarse graining operation is as a quantum channel, which is defined as a completely positive trace preserving (CPTP) map   $\Lambda_{CG}:\mc{L}(\mathcal{H}_D)\rightarrow\mc{L}(\mathcal{H}_d)$ such that $\mathrm{dim}(\mathcal{H}_D):=D > d=:\mathrm{dim}(\mathcal{H}_d)$, with $\mc{L}(\mc{H})$ representing the set of linear operators acting on $\mc{H}$~\cite{duarte,PedroM2017}.  

\subsection{A blurred and saturated detector coarse graining channel}

\begin{figure}[h]
    \centering\textit{}
    \includegraphics[width=6cm]{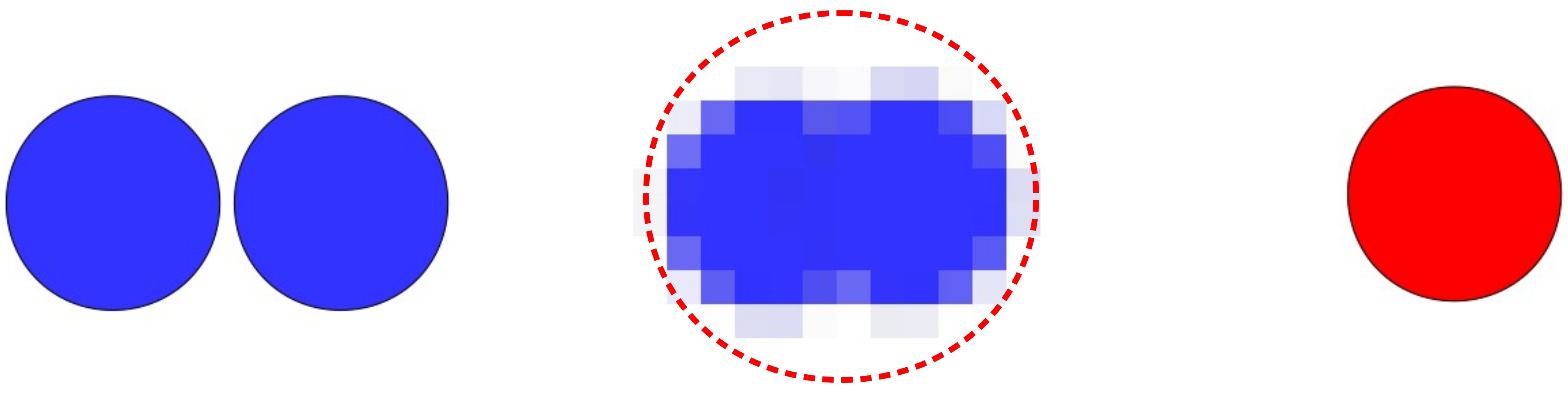}
    \caption{\textbf{Pictorial scheme of a coarse graining model.} In the left the two spheres represent a composite system of two atoms. In the middle we have a illustration suggesting the blurred view of the detection system. Then in the third picture we approximate the blurred signal as a single effective atom in a coarse-grained level.}
    \label{fig:CGscheme-1}
\end{figure}

Back to the experimental situation, imagine that we want to measure the light that comes from of a number of neighboring atoms in a lattice by fluorescence imaging \cite{sherson}, but our detecting device does not have enough resolution to identify the light coming from each individual atom. To describe this situation we can construct a coarse graining model in such way that we take the information of these multiple unresolved signals as effectively coming from one single atom in a coarse-grained level (such situation is pictorially illustrated in Fig. \ref{fig:CGscheme-1}).  

Bringing this situation to the scenario of quantum information, we can identify each atom with a qubit (a spin-$1/2$). In this way a system of  $N$ spins is described by a $D$-dimensional state $\psi\in\mc{L}(\mathcal{H}_D)$, with $D=2^N$. So we want to construct a coarse graining map $\Lambda_{CG}^{N\rightarrow1}:\mc{L}(\mathcal{H}_D)\rightarrow\mc{L}(\mathcal{H}_2)$ such that it takes a system of $N$ spins to an effective coarse-grained level of a single spin.

Starting from the simplest case, where our detector can not resolve two neighboring atoms and the amount of light coming from a single atom is already sufficient to saturate the detector, the resulting signal can be related to a single atom in a coarse-grained level.  Such situation suggests the following coarse graining map $\Lambda_{CG}^{2\rightarrow1}:\mathcal{L}(\mathcal{H}_{4})\rightarrow\mathcal{L}(\mathcal{H}_2)$, introduced in \cite{duarte,PedroM2017}:
\begin{equation}
\begin{tabular}{l|l}
$\Lambda_{CG}^{2\rightarrow1}(\ket{00}\!\bra{00})
=\ket{0}\!\bra{0}$ & $\Lambda_{CG}^{2\rightarrow1}(\ket{10}\!\bra{00})
=\frac{1}{\sqrt{3}}\ket{1}\!\bra{0}$ \\

$\Lambda_{CG}^{2\rightarrow1}(\ket{00}\!\bra{01})
=\frac{1}{\sqrt{3}}\ket{0}\!\bra{1}$ & $\Lambda_{CG}^{2\rightarrow1}(\ket{10}\!\bra{01})
=0$ \\

$\Lambda_{CG}^{2\rightarrow1}(\ket{00}\!\bra{10})
=\frac{1}{\sqrt{3}}\ket{0}\!\bra{1}$ & $\Lambda_{CG}^{2\rightarrow1}(\ket{10}\!\bra{10})
=\ket{1}\!\bra{1}$ \\

$\Lambda_{CG}^{2\rightarrow1}(\ket{00}\!\bra{11})
=\frac{1}{\sqrt{3}}\ket{0}\!\bra{1}$ & $\Lambda_{CG}^{2\rightarrow1}(\ket{10}\!\bra{11})
=0$ \\

$\Lambda_{CG}^{2\rightarrow1}(\ket{01}\!\bra{00})
=\frac{1}{\sqrt{3}}\ket{1}\!\bra{0}$ & $\Lambda_{CG}^{2\rightarrow1}(\ket{11}\!\bra{00})
=\frac{1}{\sqrt{3}}\ket{1}\!\bra{0}$ \\

$\Lambda_{CG}^{2\rightarrow1}(\ket{01}\!\bra{01})
=\ket{1}\!\bra{1}$ & $\Lambda_{CG}^{2\rightarrow1}(\ket{11}\!\bra{01})
=0$ \\

$\Lambda_{CG}^{2\rightarrow1}(\ket{01}\!\bra{10})
=0$ & $\Lambda_{CG}^{2\rightarrow1}(\ket{11}\!\bra{10})
=0$ \\

$\Lambda_{CG}^{2\rightarrow1}(\ket{01}\!\bra{11})
=0$ & $\Lambda_{CG}^{2\rightarrow1}(\ket{11}\!\bra{11})
=\ket{1}\!\bra{1}$
\label{eq:CGmap}
\end{tabular}
\end{equation}
The heuristics that lead us to construct such a map are as follows: as the detector cannot resolve if the fluorescence light comes from one atom or the other, then both states $\proj{01}$ and $\proj{10}$ lead to an effective single excitation $\proj{1}$. Moreover, we assume that the detector makes no distinction between one or two excitations, i.e. that it saturates already with a single excitation signal. As such, the state $\proj{11}$ is also mapped to the effective state $\proj{1}$. The coherence between the excited subspace, $\text{span}(\{\ket{01}, \ket{10},\ket{11}\})$, with the no-excitation subspace $\ket{00}$, maps to the effective coherence $1/\sqrt{3} |1\>\!\<0|$. The factor $1/\sqrt{3}$ comes from the dimensionality of the  subspaces, and it ensures the complete positiveness of the  coarse graining map. Lastly,  since we cannot distinguish the states $\ket{01}$, $\ket{10}$ and $\ket{11}$, there can be no coherence within this subspace, so the above null terms appear. The Hermitian-conjugated instances follow immediately from the Hermiticity preserving property of quantum channels.  It is also important to notice that this  map is not simply a partial trace, and as such it might be seen as a generalization of a decoherence process.

\begin{figure}[h]
    \centering
    \includegraphics[width=9cm]{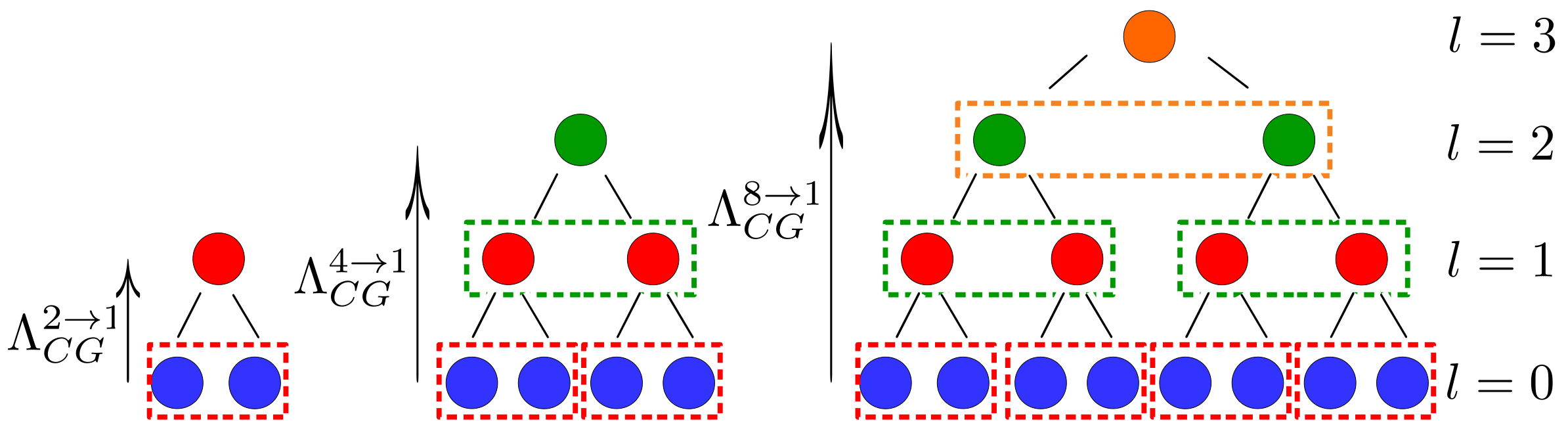}
    \caption{From a composition of $\Lambda_{CG}^{2\rightarrow1}$, higher dimensional coarse graining operations $\Lambda_{CG}^{4\rightarrow1}$  and $\Lambda_{CG}^{8\rightarrow1}$ can be defined.} 
    \label{fig:CGscheme}
\end{figure}

In the case where the detector cannot resolve more atoms, we can compose the above map as follows:
 \begin{align}
    &\Lambda_{CG}^{4\rightarrow1}=\Lambda_{CG}^{2\rightarrow1}\circ(\Lambda_{CG}^{2\rightarrow1}\otimes\Lambda_{CG}^{2\rightarrow1}), \\
    &\Lambda_{CG}^{8\rightarrow1}=\Lambda_{CG}^{2\rightarrow1}\circ(\Lambda_{CG}^{2\rightarrow1}\otimes\Lambda_{CG}^{2\rightarrow1})\circ \nonumber\\
	&\quad\quad\quad\quad\quad\quad\quad\quad\circ(\Lambda_{CG}^{2\rightarrow1}\otimes\Lambda_{CG}^{2\rightarrow1}\otimes\Lambda_{CG}^{2\rightarrow1}\otimes\Lambda_{CG}^{2\rightarrow1}), \\
	&\Lambda_{CG}^{N\rightarrow1}=\Lambda_{CG}^{2\rightarrow1}\circ(\Lambda_{CG}^{2\rightarrow1}\otimes\Lambda_{CG}^{2\rightarrow1})\circ\cdots\circ \nonumber\\
	&\quad\quad\quad\quad\quad\quad\quad\quad\circ\bigg(\bigotimes_{k=1}^{N/4}\Lambda_{CG}^{2\rightarrow1}\bigg)\circ\bigg(\bigotimes_{k=1}^{N/2}\Lambda_{CG}^{2\rightarrow1}\bigg),
\label{eq:CGcompositions}
\end{align}
where "$\circ$" denotes composition of maps. For later convenience, we define the coarse graining level $l$ as the number of times a layer of coarse graining maps is applied: for the microscopic level, where no coarse graining operation is applied we have $l=0$; starting from $N$ qubits in the microscopic level a single effective qubit will be obtained at level $l=\log N$ (we use logarithms in base 2 throughout the article) after the application of successive layers of coarse graining maps.  This process is schematically represented in Fig. \ref{fig:CGscheme}.

\section{Entanglement spreading in a Coarse-grained Spin Chain}
\label{sec:cgspinchain}

Now that we have already constructed the coarse graining map that plays the role of a blurred detector, Eq.~(\ref{eq:CGmap}), let's analyze  how entanglement evolves under this coarse graining view. In the same way as the entanglement detection was studied in \cite{fukuhara}, we will calculate the entanglement behavior between two symmetric sites around the center of a spin-chain, but now considering coarse-grained sites, that is, entanglement between two coarse-grained blocks of spins. 

Before we start, in order to properly write the microscopic ($l=0$) reduced state on the blocks $A=\{A_1,\ldots,A_N\}$ and $B=\{B_1,\ldots,B_N\}$, $\psi_{AB}^{l=0}$, it is convenient to define the following family of vectors:   
\begin{align}
\label{eq:kn}
&\ket{k_N}\!_A\equiv\ket{0_{A_1}, \cdots, 0_{A_{k-1}}, 1_{A_k}, 0_{A_{k+1}}\cdots, 0_{A_N}},  \\
&\ket{0_{N}}\!_A\equiv\ket{0_{A_1}, \cdots, 0_{A_N}},\nonumber
\end{align}
with $k\in\{1,\ldots N\}$, and similarly for $B$. It is clear that $\ket{k_N}_{A(B)}$ is a state vector with a single spin impurity at the $k$-th site of block $A(B)$, and $\ket{0_N}_{A(B)}$ is the non-impurity state. 

\subsection{$l=1:$ CG Entanglement $(2\rightarrow1)$}
\label{sec:l=1}

\begin{figure}[h!]
	\centering
	\includegraphics[width=5cm]{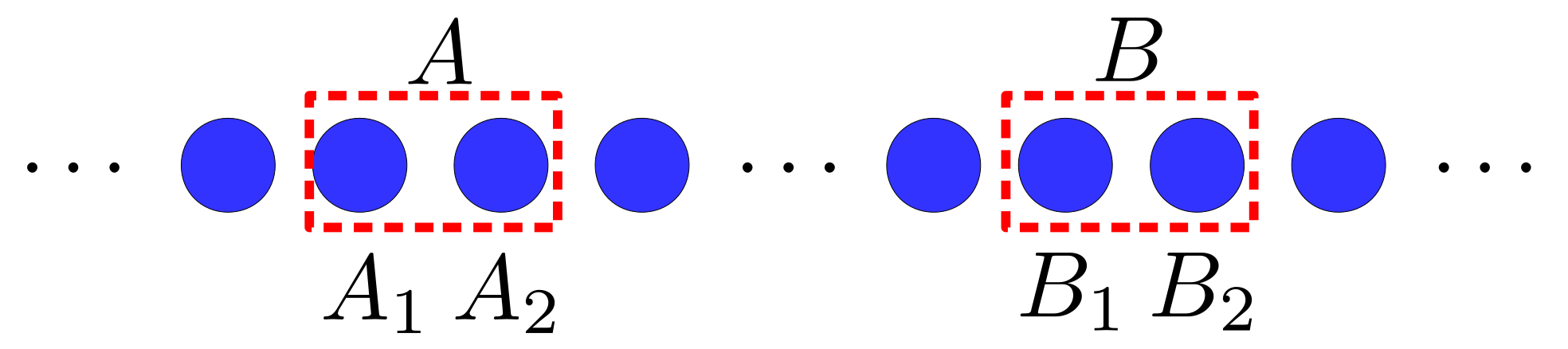}
	\caption{Schematic representation of a block of spins where the coarse graining map $\Lambda_{CG}^{2\rightarrow1}$ is going to be applied.\label{fig:CG21}}
\end{figure}

In the first situation, $l=1$, we want to compute the entanglement between two coarse-grained sites, each one coming from a block of two neighboring sites ($N=2$) at the ``microscopic'' level ($l=0$), $A=\{A_1,A_2\}$ and $B=\{B_1,B_2\}$, as is schematically represented in Fig.~\ref{fig:CG21}.
  
Remembering that we are in the single excitation subspace, and using $\{\ket{k_2}_{A(B)}\}\cup\ket{0_2}_{A(B)}$ as basis, the reduced state $\psi_{AB}^{l=0}$ can be written as
\begin{align}
&\psi_{AB}^{l=0}=\mathrm{tr}_{\overline{AB}}\ket{\psi(t)}\!\bra{\psi(t)} \nonumber \\
&=\zeta\;\ket{0_{2}}\!_A\!\bra{0_{2}}\;{\otimes}\;\ket{0_{2}}\!_B\!\bra{0_{2}}+ \nonumber\\
&+\sum_{k=1}^{2}(|\phi_{A_{k}}|^2\ket{k_{2}}\!_A\!\bra{k_{2}}\;{\otimes}\;\ket{0_{2}}\!_B\!\bra{0_{2}}+ \nonumber \\
&\quad\quad\quad\quad\quad\quad\quad\quad+|\phi_{B_{k}}|^2\ket{0_{2}}\!_A\!\bra{0_{2}}\;{\otimes}\;\ket{k_{2}}\!_B\!\bra{k_{2}})+ \nonumber\\
&+\sum_{k{\neq}k'=1}^{2}\big(\phi_{A_{k}}\phi_{A_{k'}}^\ast\ket{k_2}\!_A\!\bra{k_2'}\;{\otimes}\;\ket{0_{2}}\!_B\!\bra{0_{2}} \nonumber+ \\
&\quad\quad\quad\quad\quad\quad\quad\quad+\phi_{B_{k}}\phi_{B_{k'}}^\ast\ket{0_{2}}\!_A\!\bra{0_{2}}\;{\otimes}\;\ket{k_2}\!_B\!\bra{k_2'}\big)+ \nonumber\\
&+\sum_{k,k'=1}^{2}\big(\phi_{A_{k}}\phi_{B_{k'}}^\ast\ket{k_2}\!_A\!\bra{0_2}\;{\otimes}\;\ket{0_{2}}\!_B\!\bra{k_{2}'}+ \nonumber \\
&\quad\quad\quad\quad\quad\quad\quad\quad+\phi_{A_{k}}^\ast\phi_{B_{k'}}\ket{0_2}_A\!\bra{k_2}\;{\otimes}\;\ket{k_{2}'}_B\!\bra{0_{2}}\big).
\label{eq:redstate2}
\end{align}
with $\ket{\psi(t)}$ given by Eq.~(\ref{eq:impurityspread}), and $\zeta=1-\sum_{k=1}^2\big(|\phi_{A_k}|^2+|\phi_{B_k}|^2\big)$. In this form, it is now simple to apply the coarse graining map defined in (\ref{eq:CGmap}). Observe that the internal coherences $\phi_{A_{k}}\phi_{A_{k'}}^\ast$ and $\phi_{B_{k}}\phi_{B_{k'}}^\ast$ (for $k{\neq}k'$) related to each spin-block present in state (\ref{eq:redstate2}) no longer survive in the coarse grained state, as  $\Lambda_{CG}^{2\rightarrow1}(\ket{10}\!\bra{01})=\Lambda_{CG}^{2\rightarrow1}(\ket{01}\!\bra{10})=0$. The effective two-qubit is then:
\begin{align}
\psi^{l=1}_{AB}&=(\Lambda_{CG}^{2\rightarrow1}\otimes\Lambda_{CG}^{2\rightarrow1})(\psi_{AB}^{l=0}) \nonumber \\
&=\begin{pmatrix}
\zeta  &           0        &          0          &           0             \\
0      &       \sum\limits_{k=1}^{2}|\phi_{B_k}|^2    & \frac{1}{3}\sum\limits_{k,k'=1}^{2}\phi_{A_{k}}\phi_{B_{k'}}^\ast &           0             \\
0      & \frac{1}{3}\sum\limits_{k,k'=1}^{2}\phi_{A_{k}}^\ast\phi_{B_{k'}} &     \sum\limits_{k=1}^2|\phi_{A_k}|^2      &           0             \\
0 &            0        &          0         &  0
\end{pmatrix},
\label{eq:rhoCG}
\end{align}
where we used the basis states $\{\ket{k_1}_{A(B)}\}\cup\ket{0_1}_{A(B)}$, for each effective site.

The entanglement between two coarse-grained sites is derived analogously to that found before for two (non-coarse-grained) sites in (\ref{eq:concAB}). Since the state (\ref{eq:rhoCG}) takes the $X$-form, its concurrence is given by   
\beq
\mathcal{C}(\psi^{l=1}_{AB})=\dfrac{2}{3}\bigg|\sum_{k,k'=1}^{2}\phi_{A_k}\phi_{B_{k'}}^*\bigg|.
\label{eq:concCGABCD}
\eeq

For the purpose of giving a concrete view of the consequences of this result, we numerically investigated the entanglement evolution  in a seventeen sites chain with the spin-impurity beginning at its center. In Fig.~\ref{fig:conc12,34}(a) we plot in blue the concurrence between symmetric sites around the center of the spin-chain in the "microscopic" level, and in red the concurrence considering their related coarse-grained sites. In the microscopic level we use Eq.~(\ref{eq:concAB}) to calculate the concurrence between the first two pairs of sites before the coarse graining: $\mathcal{C}(\psi_{1,-1}^{l=0})=2|\phi_1\phi_{-1}^\ast|$ and $\mathcal{C}(\psi_{2-2}^{l=0})=2|\phi_2\phi_{-2}^\ast|$. Then, we calculate the concurrence of the resulting pair of coarse-grained sites using Eq.~(\ref{eq:concCGABCD}): $\mathcal{C}(\psi_{1,-1}^{l=1})=(2/3)|(\phi_1+\phi_2)(\phi_{-1}+\phi_{-2})^\ast|$. In the same way we calculate the concurrence between other symmetric sites around the center both at the ``microscopic'' level and at coarse-grained level with $l=1$. 

As expected, we observe that concurrence decays in the coarse graining level. Despite of this, a significant amount of entanglement still survives. So even in the coarse-grained description we observe a propagating entanglement ``wave'' (See Fig. \ref{fig:conc12,34}(b)), with values above the error bar  (black dashed line) observed by the experimental realization in~\cite{fukuhara}. 
It is interesting to notice that the maximum value of concurrence values in coarse graining level $l=1$ are in better agreement with the experimental results showed in \cite{fukuhara} than when compared with the theoretical results in ``microscopic'' level $l=0$. One can thus speculate that the experimental procedure in~\cite{fukuhara} does take into account some coarse graining as the one introduced here. 

\begin{figure}[h!]
	\begin{tabular}{rc}
		(a) & \includegraphics[width=7cm,valign=t]{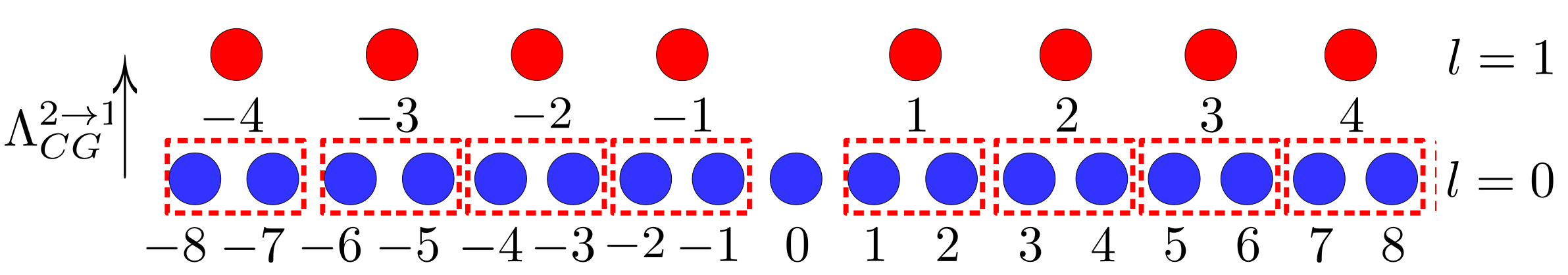} \\
		(b) & \includegraphics[width=8cm,valign=t]{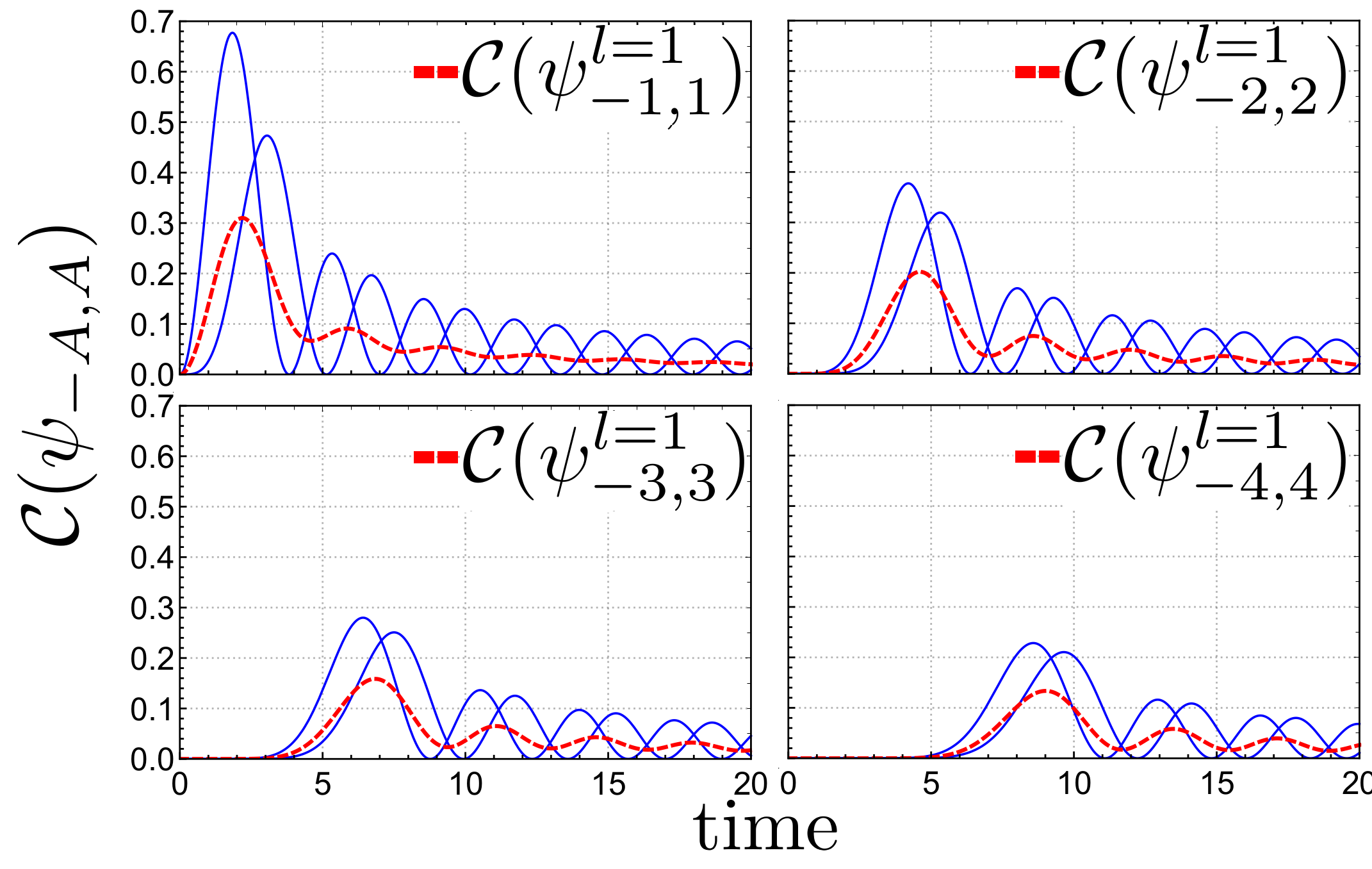} \\
		(c) & \includegraphics[width=8cm,valign=t]{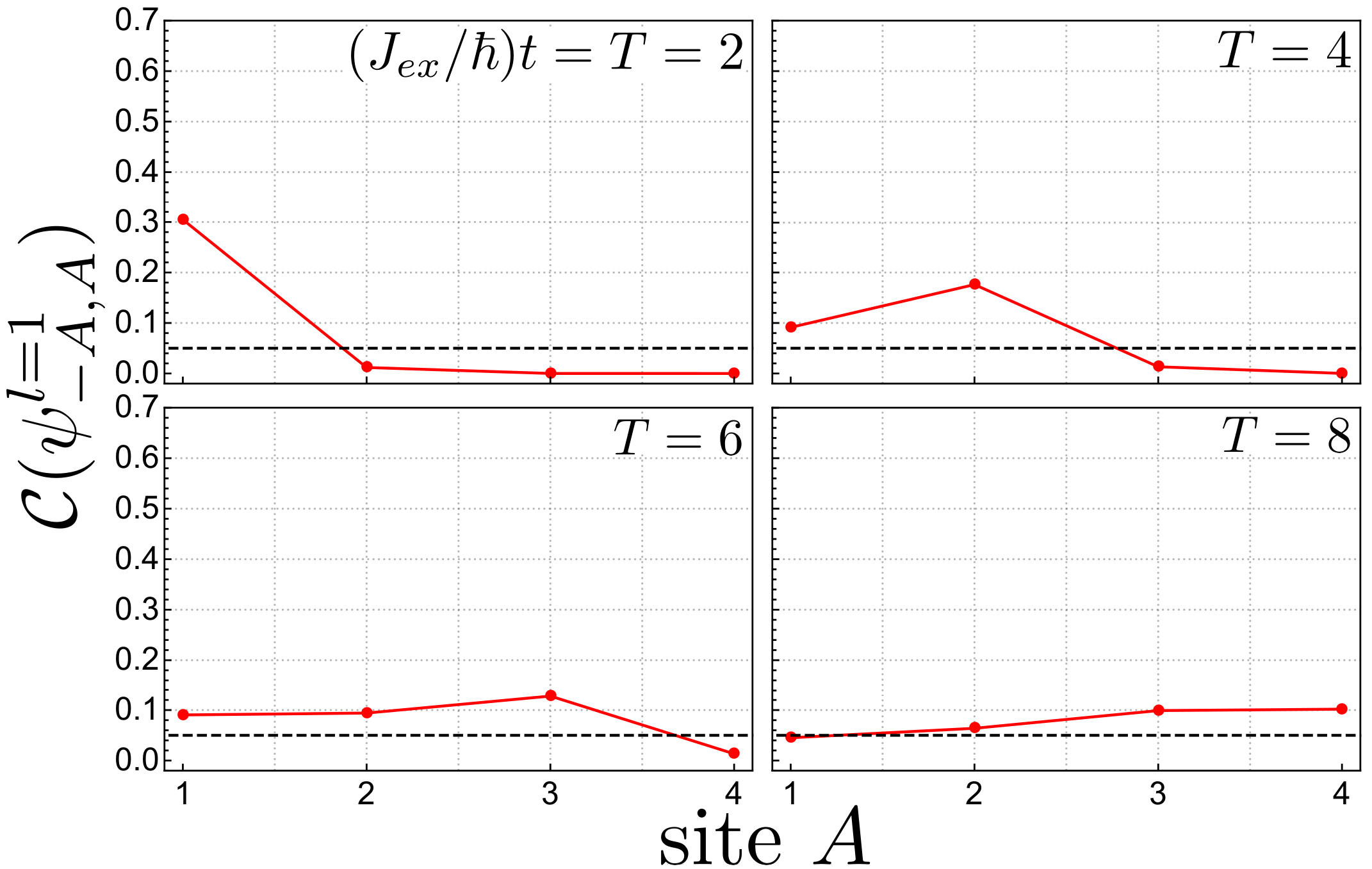} \\
		 (d) &\includegraphics[width=7cm,valign=t]{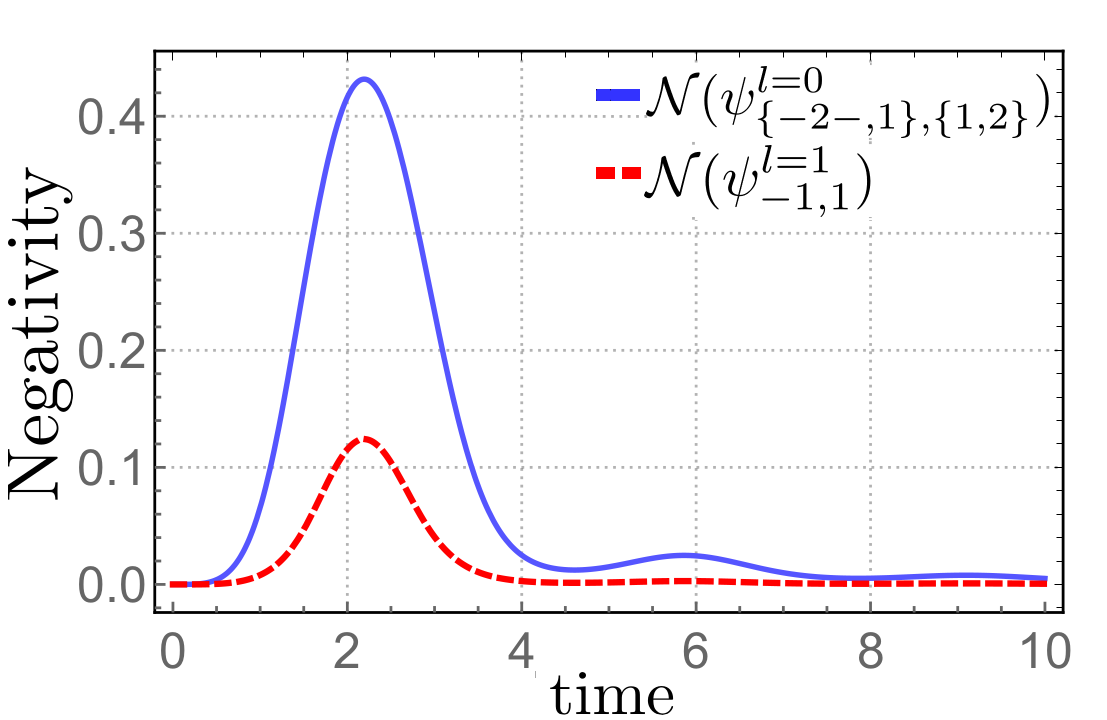}
	\end{tabular}
	\caption{\textbf{Coarse-grained entanglement in $l=1$.} (a) Effective spin chain with a single layer of coarse graining. (b) Comparison between concurrence evolution of the first four symmetric pairs of coarse-grained sites, $\mathcal{C}(\psi_{-A,A}^{l=0})$ (dashed red line), with respect to the concurrence between their relative ``microscopic'' sites (full blue lines).  (c) Entanglement wave in the coarse-grained level. The black dashed constant line represents the error in experimental detection (taken from~\cite{fukuhara}). (d) Comparison between negativity evolution of the first symmetric pairs of coarse-grained sites, $\mathcal{N}(\psi_{-1,1}^{l=1})$ (dashed red line), and the negativity $\mathcal{N}(\psi_{\{-2,-1\},\{1,2\}}^{l=0})$ among their relative ``microscopic'' sites (full blue line). The time evolution is given in dimensionless unit $J_{ex}t/\hbar$.}
	\label{fig:conc12,34}
\end{figure}

A complementary way of studying the differences in entanglement between the ``microscopic'' level and the coarse-grained level is through negativity~\cite{Neg2002}. Using concurrence we are somewhat restricted to calculate only bipartite entanglement for systems with two spins (two sites in the spin-chain). Negativity, on the other hand, can be evaluated for bipartite systems of any local dimensions. 
Here we restrict our analysis to the coarse-grained state $\psi_{-1,1}^{l=1}$ and its related ``microscopic'' state $\psi_{\{-2,-1\},\{1,2\}}^{l=0}$.  
Results are shown in Fig.\ref{fig:conc12,34} (d). Differently from the concurrence approach, with negativity we can properly compare the entanglement within blocks in different ranges of description. We observe that entanglement in the coarse graining level is smaller than the \emph{total} entanglement in the ``microscopic'' state. 

\subsection{$l=2:$ CG Entanglement ($4\rightarrow1)$}
\label{sec:l=2}

\begin{figure}[h!]
	\centering
	\includegraphics[width=7cm]{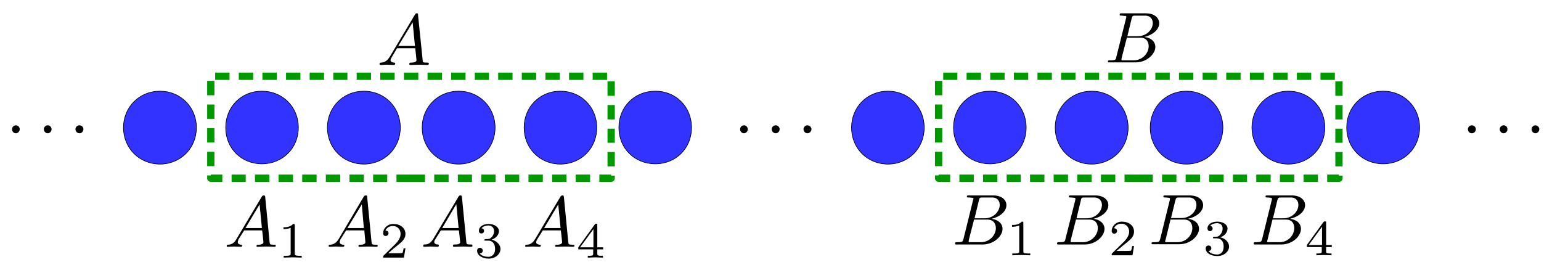}
	\caption{Schematic representation of a block of spins where the coarse graining map $\Lambda_{CG}^{4\rightarrow1}$ is going to be applied.\label{fig:CG41}}
\end{figure}

For the next level of resolution ($l=2$), we map four sites ($N=4$) as a single effective one. Given an arbitrary pair of spin-blocks composed by four neighboring atoms $A=\{A_4,A_3,A_2,A_1\}$ and $B=\{B_1,B_2,B_3,B_4\}$-- as shown in Fig.~\ref{fig:CG41} -- we proceed in the same way as above to find the coarse-grained state:
\begin{align}
\psi^{l=2}_{AB}&=(\Lambda_{CG}^{4\rightarrow1}\otimes\Lambda_{CG}^{4\rightarrow1})(\psi_{AB}^{l=0}) \nonumber \\
&=\begin{pmatrix}
\zeta  &           0        &          0          &           0             \\
0      &       \sum\limits_{k=1}^{4}|\phi_{B_k}|^2    & \frac{1}{3^2}\sum\limits_{k,k'=1}^{4}\phi_{A_{k}}\phi_{B_{k'}}^\ast &           0             \\
0      & \frac{1}{3^2}\sum\limits_{k,k'=1}^{4}\phi_{A_{k}}^\ast\phi_{B_{k'}} &     \sum\limits_{k=1}^4|\phi_{A_k}|^2      &           0             \\
0 &            0        &          0         &  0
\end{pmatrix}.
\label{eq:rhoCG41}
\end{align}

As observed in the previous subsection (\ref{sec:l=1}), the internal coherence related to spin-blocks, $\phi_{A_{k}}\phi_{A_{k'}}^\ast$ and $\phi_{B_{k}}\phi_{B_{k'}}^\ast$, with $k\neq k'$, do not  survive in the effective state (\ref{eq:rhoCG41}). As before, this happens because  $\Lambda_{CG}^{2\rightarrow1}(\ket{10}\!\bra{01})=\Lambda_{CG}^{2\rightarrow1}(\ket{01}\!\bra{10})=0$, and due to the fact that $\Lambda_{CG}^{4\rightarrow1}=\Lambda_{CG}^{2\rightarrow1}\circ(\Lambda_{CG}^{2\rightarrow1}\otimes\Lambda_{CG}^{2\rightarrow1})$. The square power on the $1/3$ factor reflects the two layers of coarse graining which were applied.

For such a state the concurrence can be easily evaluated to give:
\beq
\mathcal{C}(\psi^{l=2}_{AB})=\dfrac{2}{3^2}\bigg|\sum_{k,k'=1}^{4}2\phi_{A_k}\phi_{B_{k'}}^*\bigg|.
\label{eq:concCG41}
\eeq
Again we analyze the entanglement evolution by looking at the seventeen centralized sites of the ``microscopic'' chain ($l=0$) whose spin-impurity begins at its center. 
Results are shown in Fig.\ref{fig:conc1234,5678}.
We use equation (\ref{eq:concAB}) to calculate the concurrence between the first four pair of sites before the coarse graining (in blue): $\mathcal{C}(\psi_{-1,1}^{l=0})\!=\!2|\phi_{-1}\phi_{1}^\ast|$, $\mathcal{C}(\psi_{-2,2}^{l=0})\!=\!2|\phi_{-2}\phi_{2}^\ast|$, $\mathcal{C}(\psi_{-2,3}^{l=0})\!=\!2|\phi_{-3}\phi_{3}^\ast|$ and $\mathcal{C}(\psi_{-4,4}^{l=0})\!=\!2|\phi_{-4}\phi_{4}^\ast|$. Then, we calculate the concurrence of resulting pair of coarse-grained sites using the equation \eqref{eq:concCG41} (in green): $\mathcal{C}(\psi_{-1,1}^{l=2})\!=\!(2/9)|(\phi_{-1}\!+\!\phi_{-2}\!+\!\phi_{-3}\!+\!\phi_{-4})(\phi_{1}\!+\!\phi_{2}\!+\!\phi_{3}\!+\!\phi_{4})^{\ast}|$. We proceed in the same way to calculate the concurrence between the other symmetric sites. 

As expected, we observe that the concurrence becomes weaker when compared to the concurrence in the ``microscopic'' level or even with the one in the first coarse grained level (\ref{fig:conc12,34}). Consequently we observe an very weak entanglement ``wave'', in the limit of the experimental error detection (see \ref{fig:conc12,34}(c)). Therefore in this scheme, if we consider the error bars of the experimental work \cite{fukuhara}, our results suggest that entanglement would no longer be detectable.

\begin{figure}[h!]	
	\begin{tabular}{lr}
	(a)	&\includegraphics[width=7cm,valign=t]{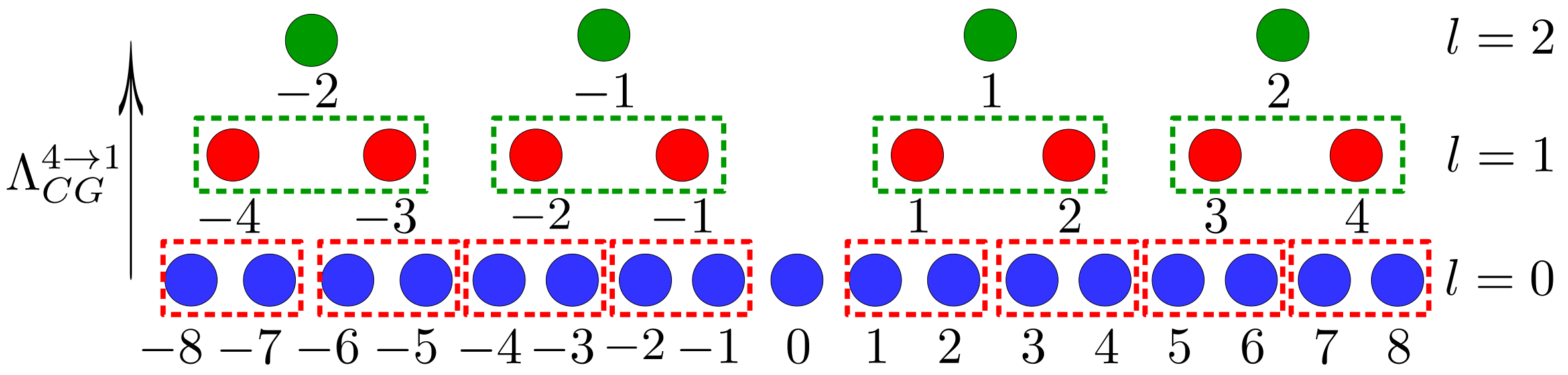} \\
	(b)	&\includegraphics[width=8cm,valign=t]{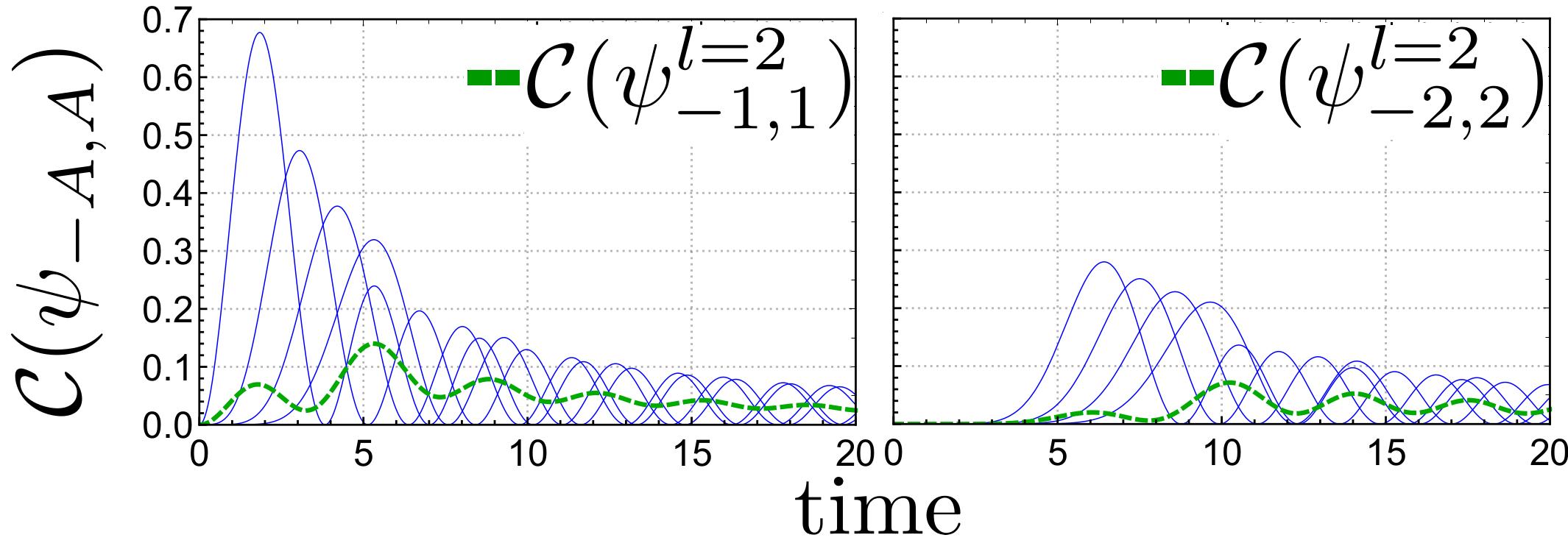}   \\
	(c)	&\includegraphics[width=8cm,valign=t]{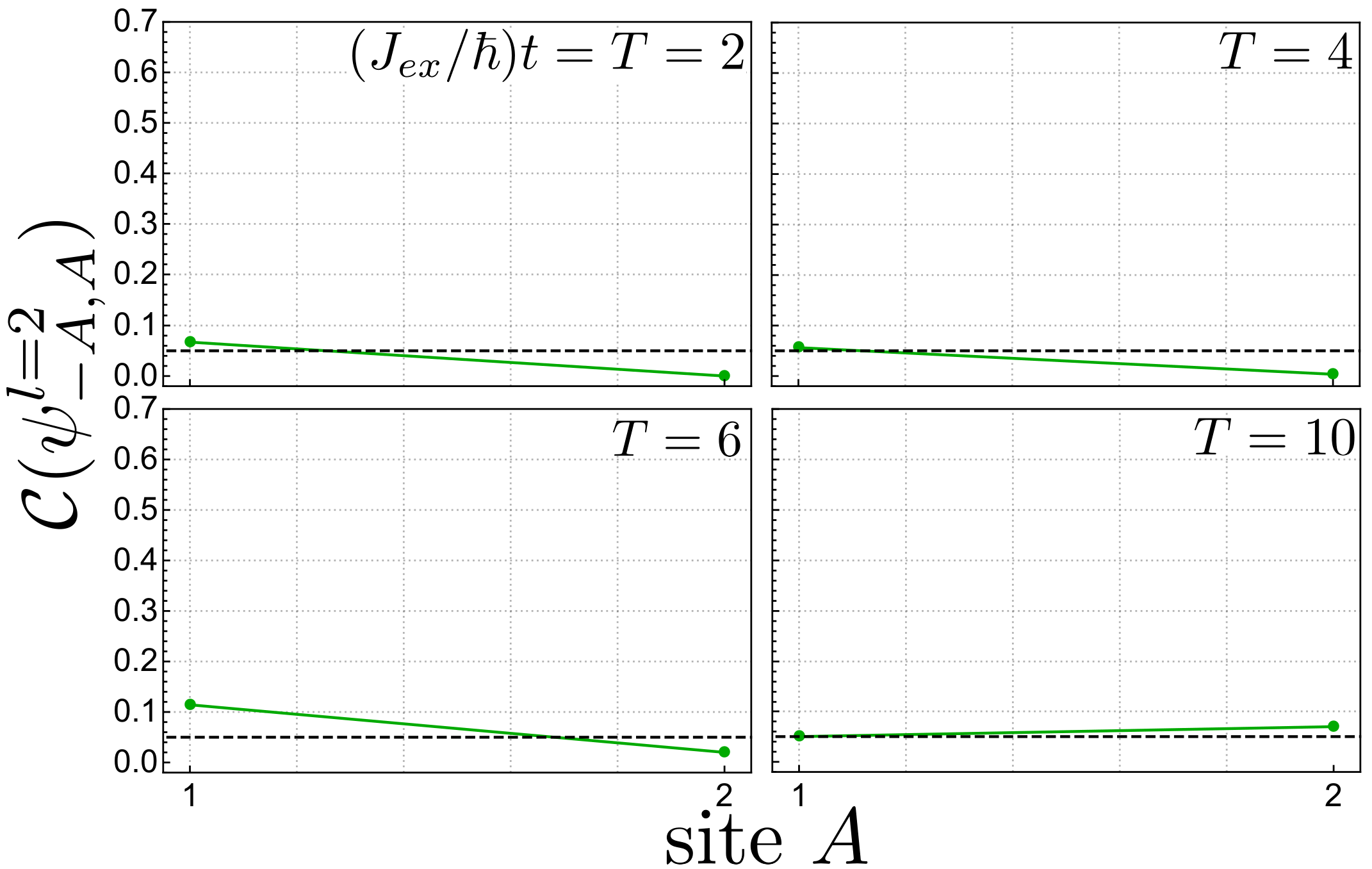} \\
	(d)     &\includegraphics[width=7cm,valign=t]{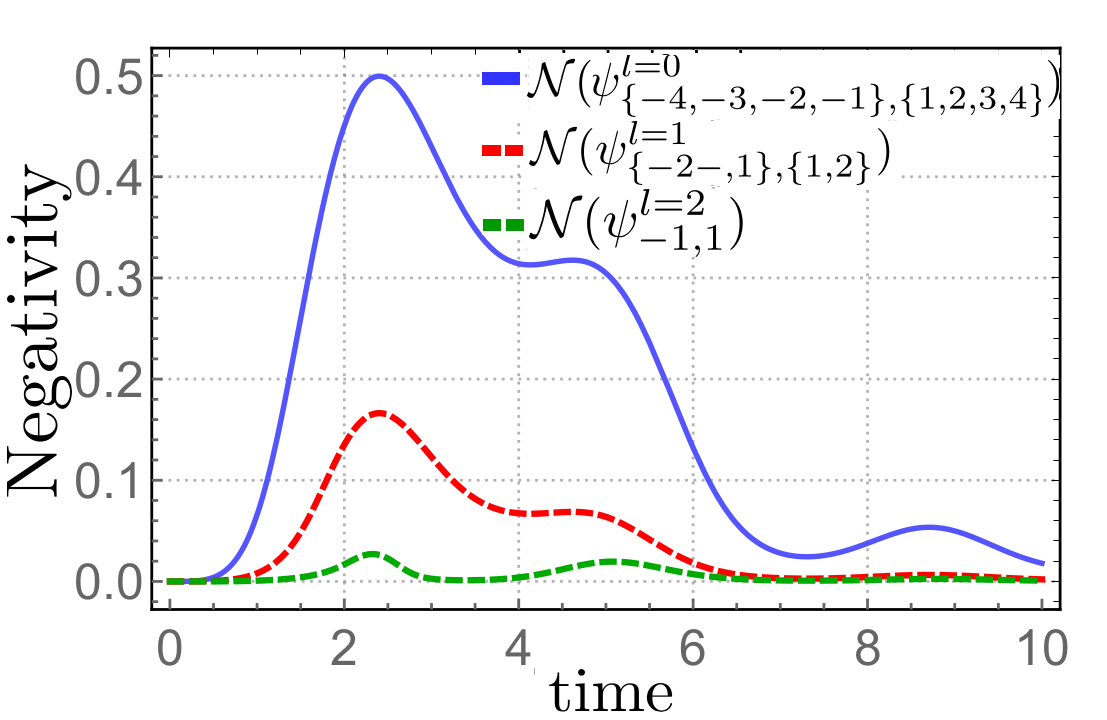}   
    \end{tabular}
	\caption{\textbf{Coarse-grained entanglement in $l=2$.} (a) Effective spin-chain with two layers of coarse graining. (b) Comparison between concurrence evolution of the first two symmetric pairs of coarse-grained sites, $\mathcal{C}(\psi_{-A,A}^{l=2})$ (dashed green line), and the concurrence between their relative ``microscopic''-sites (full blue lines). The time evolution is given in dimensionless unit $J_{ex}t/\hbar$ (c) Spatial dynamics of concurrence in the CG level. The black dashed constant line represents the error in experimental detection. (d) Comparison between negativity evolution in different levels of description, $l=1$ (upper dashed red line), $l=2$ (lower dashed green line), and the negativity in the microscopic supercell, $l=0$ (full blue line). The time evolution is given in dimensionless unit $J_{ex}t/\hbar$.}
	\label{fig:conc1234,5678}
\end{figure}

Analogously to the last subsection, here we calculate the negativity for the different description levels of the system $l=0,1,2$. We consider the coarse-grained states  $\psi_{-1,1}^{l=2}$ and $\psi_{\{-2,-1\},\{1,2\}}^{l=1}$, and their correspondent ``microscopic'' state  $\psi_{\{-4,-3,-2,-1\},\{1,2,3,4\}}^{l=0}$.
 Results are shown in Fig.~\ref{fig:conc1234,5678} (c). 
 We observe how drastic it is the effect of the coarse graining on the entanglement among different levels.
This result is in agreement with the ones for concurrence, indicating a weak entanglement signal after two layers of coarse graining (level $l=2$). That is, if the detector cannot resolve the signal coming for a ensemble of four neighboring sites in a spin-chain, entanglement will be hardly detectable in the experimental situation. Again we observe that the effective entanglement is smaller than the total entanglement in the super-cell.

\subsection{$l=\log N$: CG Entanglement ($N\rightarrow1$)}

In this section we generalize the previous concurrence results to any level of coarse graining. At the microscopic level, we consider two blocks of sites $A=\{A_1,\cdots,A_2,A_N\}$ and $B=\{B_1,B_2,\cdots,B_N\}$, in which we will apply $\log N$ layers of the coarse graining map:
\begin{align}
&\Lambda_{CG}^{N\rightarrow1}=\Lambda_{CG}^{2\rightarrow1}\circ(\Lambda_{CG}^{2\rightarrow1}\otimes\Lambda_{CG}^{2\rightarrow1})\circ\cdots\circ \nonumber\\
&\quad\quad\quad\quad\quad\quad\quad\quad\circ\bigg(\bigotimes_{k=1}^{N/4}\Lambda_{CG}^{2\rightarrow1}\bigg)\circ\bigg(\bigotimes_{k=1}^{N/2}\Lambda_{CG}^{2\rightarrow1}\bigg).
\end{align}

Using the total pure state in equation \eqref{eq:impurityspread}, and definition (\ref{eq:kn}), we can write the reduced state over the blocks $A$ and $B$ as:
\begin{align}
\psi_{AB}^{l=0}&=\zeta\;\ket{0_{N}}\!_A\!\bra{0_{N}}\;{\otimes}\;\ket{0_{N}}\!_B\!\bra{0_{N}}+ \nonumber\\
&+\sum_{k=1}^{N}(|\phi_{A_{k}}|^2\ket{k_{N}}\!_A\!\bra{k_{N}}\;{\otimes}\;\ket{0_{N}}\!_B\!\bra{0_{N}}+ \nonumber \\
&\quad\quad\quad\quad\quad\quad+|\phi_{B_{k}}|^2\ket{0_{N}}\!_A\!\bra{0_{N}}\;{\otimes}\;\ket{k_{N}}\!_B\!\bra{k_{N}})+ \nonumber\\
&+\sum_{k{\neq}k'=1}^{N}\big(\phi_{A_{k}}\phi_{A_{k'}}^\ast\ket{k_N}\!_A\!\bra{k_N'}\;{\otimes}\;\ket{0_{N}}\!_B\!\bra{0_{N}} \nonumber+ \\
&\quad\quad\quad\quad\quad\quad+\phi_{B_{k}}\phi_{B_{k'}}^\ast\ket{0_{N}}\!_A\!\bra{0_{N}}\;{\otimes}\;\ket{k_N}\!_B\!\bra{k_N'}\big)+ \nonumber\\
&+\sum_{k,k'=1}^{N}\big(\phi_{A_{k}}\phi_{B_{k'}}^\ast\ket{k_N}\!_A\!\bra{0_N}\;{\otimes}\;\ket{0_{N}}\!_B\!\bra{k_{N}'}+ \nonumber \\
&\quad\quad\quad\quad\quad\quad+\phi_{A_{k}}^\ast\phi_{B_{k'}}\ket{0_N}\!_A\!\bra{k_N}\;{\otimes}\;\ket{k_{N}'}\!_B\!\bra{0_{N}}\big),
\label{eq:redstateN}
\end{align}
where $\zeta=1-\sum_{k=1}^N\big(|\phi_{A_k}|^2+|\phi_{B_k}|^2\big)$.

The rational to apply the coarse graining map in both supercells,  $\Lambda_{CG}^{N\rightarrow1}\otimes \Lambda_{CG}^{N\rightarrow1}$, is as before. The first term contains no excitations, and as such goes to $\zeta\ket{0}\!_A\!\bra{0}\otimes \ket{0}\!_B\!\bra{0}$. The second term contains the diagonal/population elements where the excitation is either in part $A$ or in part $B$. As the coarse graining maps states in the single-excitation subspace into (smaller dimensional) single-excitation subspace, this term goes to $\sum_{k=1}^{N}(|\phi_{A_{k}}|^2\ket{1}\!_A\!\bra{1}\otimes\ket{0}\!_B\!\bra{0} + |\phi_{B_{k}}|^2\ket{0}\!_A\!\bra{0}\otimes\ket{1}\!_B\!\bra{1})$. The third term contains off-diagonal/coherence elements where the excitation is either in the part $A$ or in the part $B$. As $k\neq k'$, these elements represent coherences between states that the detector cannot distinguish, and as such, at some level, these elements vanish. Lastly we have the term with coherences between the single-excitation subspace and the no-excitation subspace within a supercell. At a given layer there will always be per block one element  of the form $\ket{01}\!_A\!\bra{00}$ or $\ket{10}\!_A\!\bra{00}$, with all others being  as $\ket{00}\!_A\!\bra{00}$. The action of the coarse graining map on the first type of elements gives $\ket{1}\!_A\!\bra{0}/\sqrt{3}$, while on the second gives $\ket{0}\!_A\!\bra{0}$. The same reasoning applies for the Hermitian conjugate elements and for the block $B$. Since we start with $N$ spins per block, we can repeat this process $\log N$ times, and as such, at level $l=\log N$, this term goes to $(1/3)^{\log N} \sum_{k,k'=1}^{N}\phi_{A_{k}}\phi_{B_{k'}}^\ast\ket{1}\!_A\!\bra{0}\otimes\ket{0}\!_B\!\bra{1}+\phi_{A_{k}}^\ast\phi_{B_{k'}}\ket{0}\!_A\!\bra{1}\otimes\ket{1}\!_B\!\bra{0}$. In summary, after applying the coarse graining map $\log N$ times on the pure state  \eqref{eq:impurityspread}, we get:
\begin{align}
&\psi^{l=\log N}_{AB}=(\Lambda_{CG}^{N\rightarrow1}\otimes\Lambda_{CG}^{N\rightarrow1})(\psi_{AB}^{l=0}) \nonumber \\
&=\begin{pmatrix}
\zeta  &           0        &          0          &           0             \\
0      &       \sum\limits_{k=1}^{N}|\phi_{B_k}|^2    & \frac{1}{3^{\log N}}\sum\limits_{k,k'=1}^{N}\phi_{A_{k}}\phi_{B_{k'}}^\ast &           0             \\
0      & \frac{1}{3^{\log N}}\sum\limits_{k,k'=1}^{N}\phi_{A_{k}}^\ast\phi_{B_{k'}} &     \sum\limits_{k=1}^N|\phi_{A_k}|^2      &           0             \\
0 &            0        &          0         &  0
\end{pmatrix}.
\label{eq:redstateNCG}
\end{align}

\begin{figure}[h!]
	\includegraphics[width=7cm]{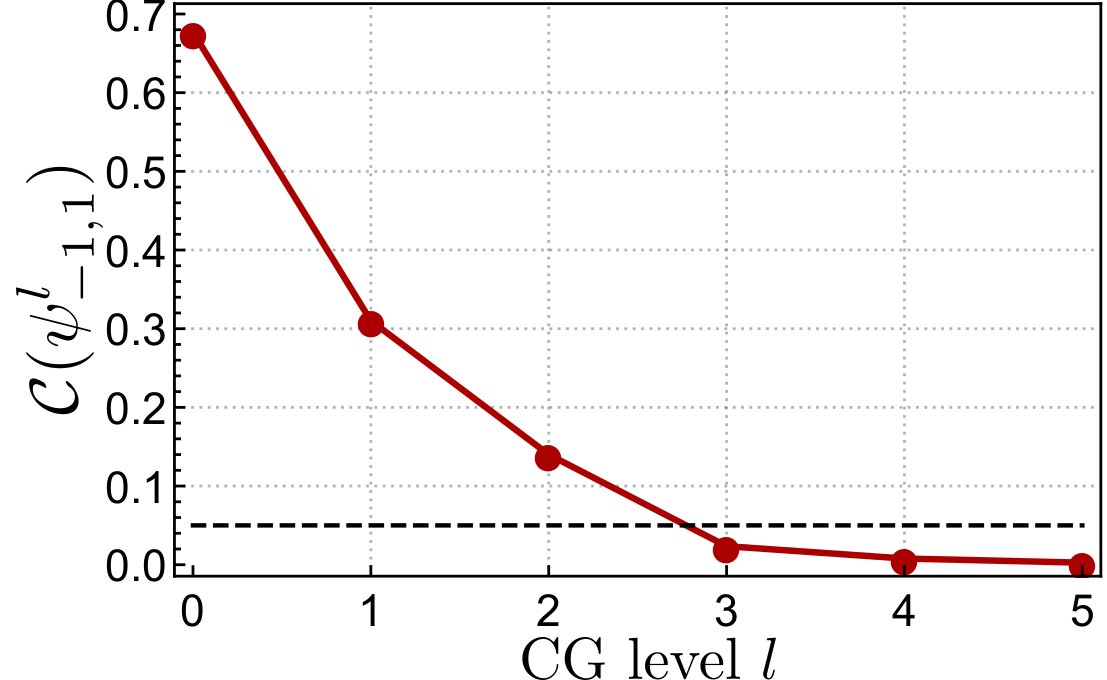}
	\caption{\textbf{Concurrence behavior at each coarse graining level.} Each dot represents the maximum value attained for the concurrence at the respective coarse graining level. The black dashed constant line represents the error in experimental detection.}
	\label{fig:ConcurrenceCGN1}
\end{figure}

As before, now it is simple to evaluate the effective concurrence:
\beq
	\mathcal{C}(\psi^{l=\log N}_{AB})=\dfrac{2}{3^{\log N}}\Big|\sum_{k,k'=1}^N\phi_{A_k}\phi_{B_{k'}}^{\ast}\Big|. 
\label{eq:concN1}
\eeq
Now we can evaluate the limit of effective concurrence detection. To do this we compare the maximum value attained for concurrence $\mathcal{C}(\psi^{l}_{-1,1})$ at each coarse graining level. Results are shown in Fig.~\ref{fig:ConcurrenceCGN1}. We note that from $l=3$ onward the values became smaller than the experimental detection error observed in~\cite{fukuhara}, which suggests that no entanglement could be detected in such a experimental resolution.    

From Eq.~\eqref{eq:concN1} it is simple to established that the effective concurrence at a given coarse graining level $l$ is smaller than the sum of all bipartite microscopic concurrences:
\beq
\mathcal{C}(\psi^{l=\log N}_{AB}) \le \dfrac{1}{3^{\log N}}\sum_{k,k'=1}^N\mathcal{C}(\psi^{l=0}_{A_{k}B_{k'}}).
\label{eq:ineqconcN1}
\eeq
We observe an exponential entanglement decay with respect to the coarse graining level $l=\log N$.

Moreover, now we can address the following physically motivated question: From the coarse-grained concurrence can we say something about entanglement at the ``microscopic'' level? Let $\mc{C}_{\text{max}}^{l=0} = \max_{k,k^\prime} \mathcal{C}(\psi^{l=0}_{A_kB_{k'}})$. Then from the above inequality we have:
\beq
\mathcal{C}_{\text{max}}^{l=0}\;{\geq}\;\bigg(\dfrac{3}{4}\bigg)^{\log N}\mathcal{C}(\psi^{l=\log N}_{AB}).
\label{eq:ineqcmax}
\eeq
Therefore with this procedure we can construct a lower bound for the maximal concurrence between a  pair of spins in the ``microscopic'' level.  Naturally, we are not able to identify the pair of spins which shares the maximum entanglement at a given time, but we can say that there is a pair of spins in the ``microscopic''  chain with entanglement at least $(3/4)^{\log{N}}$ of the concurrence measured in the ``macroscopic level''.

\section{Summary and Conclusions}
\label{sec:conclusions}

In the present work we have explored a suitable coarse graining channel, proposed in \cite{duarte,PedroM2017}, as a tool to describe entanglement spreading in a coarse-grained spin-chain with different degrees of resolution. Comparing with the experimental realizations performed with ultracold atoms in optical latices~\cite{fukuhara,fukuhara2}, our results suggest that even if we are not able to fully resolve the system, entanglement can be still detected at some coarse graining levels. With our approach we showed that even if a detector cannot resolve the signal that comes from two or four neighboring qubits an entanglement wave can be still detected. Furthermore, we showed that it is possible to have some information about the “microscopic” entanglement even if we have access only to the coarse graining description~(\ref{eq:ineqcmax}). 

Beyond that experimental situation, ultracold atoms in optical lattice represent an excellent platform for simulation of many other quantum many body problems~\cite{gross2017quantum,atomdet}. Part of this success is due to the development of high precision detection devices like quantum gas microscopes~\cite{sherson,gross2017quantum}. Since our coarse graining is constructed to model this kind of detection, we hope that our coarse graining model will be relevant for other ultracold atom-based quantum simulations. Some possibilities  are: to probe quantum magnetism, to realize and detect topological matter and the study of quantum systems with long-range interactions~\cite{gross2017quantum,atomdet}. Situations where more than one excitation are present in the system, and we have the full XXZ Hamiltonian like in Refs.~\cite{referee1,referee2},  can also be directly model by our map. Moreover, given its simplicity and intuitive construction, we expect that this coarse graining approach will be adapted and extended to other experimental achievements in which detection cannot be made with proper accuracy.

Other coarse graining situations  can also be explored. Beyond modeling a blurred detection, the coarse graining approach can be a tool to explore other problems in physics. From our results and other recent works, like \cite{duarte}, we envisage that coarse graining maps will play an important role in exploring foundations of statistical mechanics and emergent phenomenons in quantum physics. As discussed in the beginning of this paper, we hope that this approach may shed some light on the classical-to-quantum transition problem, bringing to the discussion other mechanisms that perturbs quantum resources beyond the decoherence.

\begin{acknowledgments}
	We would like to acknowledge fruitful discussions with Carlos Pineda, Fred Brito, Marcelo França, and Pedro Costa. This work is supported by the Brazilian funding agencies  CNPq and CAPES, and it is part of the Brazilian National Institute for Quantum Information.
\end{acknowledgments}

\bibliographystyle{apsrev}

\begin{thebibliography}{29}
	\expandafter\ifx\csname natexlab\endcsname\relax\def\natexlab#1{#1}\fi
	\expandafter\ifx\csname bibnamefont\endcsname\relax
	\def\bibnamefont#1{#1}\fi
	\expandafter\ifx\csname bibfnamefont\endcsname\relax
	\def\bibfnamefont#1{#1}\fi
	\expandafter\ifx\csname citenamefont\endcsname\relax
	\def\citenamefont#1{#1}\fi
	\expandafter\ifx\csname url\endcsname\relax
	\def\url#1{\texttt{#1}}\fi
	\expandafter\ifx\csname urlprefix\endcsname\relax\def\urlprefix{URL }\fi
	\providecommand{\bibinfo}[2]{#2}
	\providecommand{\eprint}[2][]{\url{#2}}
	
	\bibitem[{\citenamefont{Haroche}(2013)}]{haroche}
	\bibinfo{author}{\bibfnamefont{S.}~\bibnamefont{Haroche}},
	\bibinfo{journal}{Rev. Mod. Phys.} \textbf{\bibinfo{volume}{85}},
	\bibinfo{pages}{1083} (\bibinfo{year}{2013}),
	\urlprefix\url{https://link.aps.org/doi/10.1103/RevModPhys.85.1083}.
	
	\bibitem[{\citenamefont{Wineland}(2013)}]{wineland}
	\bibinfo{author}{\bibfnamefont{D.~J.} \bibnamefont{Wineland}},
	\bibinfo{journal}{Rev. Mod. Phys.} \textbf{\bibinfo{volume}{85}},
	\bibinfo{pages}{1103} (\bibinfo{year}{2013}),
	\urlprefix\url{https://link.aps.org/doi/10.1103/RevModPhys.85.1103}.
	
	\bibitem[{\citenamefont{Barends et~al.}(2016)\citenamefont{Barends, Shabani,
			Lamata, Kelly, Mezzacapo, Las~Heras, Babbush, Fowler, Campbell, Chen
			et~al.}}]{google}
	\bibinfo{author}{\bibfnamefont{R.}~\bibnamefont{Barends}},
	\bibinfo{author}{\bibfnamefont{A.}~\bibnamefont{Shabani}},
	\bibinfo{author}{\bibfnamefont{L.}~\bibnamefont{Lamata}},
	\bibinfo{author}{\bibfnamefont{J.}~\bibnamefont{Kelly}},
	\bibinfo{author}{\bibfnamefont{A.}~\bibnamefont{Mezzacapo}},
	\bibinfo{author}{\bibfnamefont{U.}~\bibnamefont{Las~Heras}},
	\bibinfo{author}{\bibfnamefont{R.}~\bibnamefont{Babbush}},
	\bibinfo{author}{\bibfnamefont{A.~G.} \bibnamefont{Fowler}},
	\bibinfo{author}{\bibfnamefont{B.}~\bibnamefont{Campbell}},
	\bibinfo{author}{\bibfnamefont{Y.}~\bibnamefont{Chen}}, \bibnamefont{et~al.},
	\bibinfo{journal}{Nature} \textbf{\bibinfo{volume}{534}},
	\bibinfo{pages}{222} (\bibinfo{year}{2016}),
	\urlprefix\url{https://doi.org/10.1038/nature17658}.
	
	\bibitem[{\citenamefont{Gambetta et~al.}(2017)\citenamefont{Gambetta, Chow, and
			Steffen}}]{ibm}
	\bibinfo{author}{\bibfnamefont{J.~M.} \bibnamefont{Gambetta}},
	\bibinfo{author}{\bibfnamefont{J.~M.} \bibnamefont{Chow}}, \bibnamefont{and}
	\bibinfo{author}{\bibfnamefont{M.}~\bibnamefont{Steffen}},
	\bibinfo{journal}{npj Quantum Information} \textbf{\bibinfo{volume}{3}},
	\bibinfo{pages}{2} (\bibinfo{year}{2017}),
	\urlprefix\url{https://doi.org/10.1038/s41534-016-0004-0}.
	
	\bibitem[{\citenamefont{Johnson et~al.}(2011)\citenamefont{Johnson, Amin,
			Gildert, Lanting, Hamze, Dickson, Harris, Berkley, Johansson, Bunyk
			et~al.}}]{dwave}
	\bibinfo{author}{\bibfnamefont{M.~W.} \bibnamefont{Johnson}},
	\bibinfo{author}{\bibfnamefont{M.~H.} \bibnamefont{Amin}},
	\bibinfo{author}{\bibfnamefont{S.}~\bibnamefont{Gildert}},
	\bibinfo{author}{\bibfnamefont{T.}~\bibnamefont{Lanting}},
	\bibinfo{author}{\bibfnamefont{F.}~\bibnamefont{Hamze}},
	\bibinfo{author}{\bibfnamefont{N.}~\bibnamefont{Dickson}},
	\bibinfo{author}{\bibfnamefont{R.}~\bibnamefont{Harris}},
	\bibinfo{author}{\bibfnamefont{A.~J.} \bibnamefont{Berkley}},
	\bibinfo{author}{\bibfnamefont{J.}~\bibnamefont{Johansson}},
	\bibinfo{author}{\bibfnamefont{P.}~\bibnamefont{Bunyk}},
	\bibnamefont{et~al.}, \bibinfo{journal}{Nature}
	\textbf{\bibinfo{volume}{473}}, \bibinfo{pages}{194} (\bibinfo{year}{2011}),
	\urlprefix\url{https://doi.org/10.1038/nature10012}.
	
	\bibitem[{\citenamefont{H{\"a}ffner et~al.}(2005)\citenamefont{H{\"a}ffner,
			H{\"a}nsel, Roos, Benhelm, Chwalla, K{\"o}rber, Rapol, Riebe, Schmidt, Becher
			et~al.}}]{haffner2005scalable}
	\bibinfo{author}{\bibfnamefont{H.}~\bibnamefont{H{\"a}ffner}},
	\bibinfo{author}{\bibfnamefont{W.}~\bibnamefont{H{\"a}nsel}},
	\bibinfo{author}{\bibfnamefont{C.}~\bibnamefont{Roos}},
	\bibinfo{author}{\bibfnamefont{J.}~\bibnamefont{Benhelm}},
	\bibinfo{author}{\bibfnamefont{M.}~\bibnamefont{Chwalla}},
	\bibinfo{author}{\bibfnamefont{T.}~\bibnamefont{K{\"o}rber}},
	\bibinfo{author}{\bibfnamefont{U.}~\bibnamefont{Rapol}},
	\bibinfo{author}{\bibfnamefont{M.}~\bibnamefont{Riebe}},
	\bibinfo{author}{\bibfnamefont{P.}~\bibnamefont{Schmidt}},
	\bibinfo{author}{\bibfnamefont{C.}~\bibnamefont{Becher}},
	\bibnamefont{et~al.}, \bibinfo{journal}{Nature}
	\textbf{\bibinfo{volume}{438}}, \bibinfo{pages}{643} (\bibinfo{year}{2005}),
	\urlprefix\url{https://doi.org/10.1038/nature04279}.
	
	\bibitem[{\citenamefont{Aolita et~al.}(2015)\citenamefont{Aolita, de~Melo, and
			Davidovich}}]{aolita}
	\bibinfo{author}{\bibfnamefont{L.}~\bibnamefont{Aolita}},
	\bibinfo{author}{\bibfnamefont{F.}~\bibnamefont{de~Melo}}, \bibnamefont{and}
	\bibinfo{author}{\bibfnamefont{L.}~\bibnamefont{Davidovich}},
	\bibinfo{journal}{Reports on Progress in Physics}
	\textbf{\bibinfo{volume}{78}}, \bibinfo{pages}{042001}
	(\bibinfo{year}{2015}),
	\urlprefix\url{https://doi.org/10.1088%2F0034-4885%2F78%2F4%2F042001}.
		
		\bibitem[{\citenamefont{Gross and Bloch}(2017)}]{gross2017quantum}
		\bibinfo{author}{\bibfnamefont{C.}~\bibnamefont{Gross}} \bibnamefont{and}
		\bibinfo{author}{\bibfnamefont{I.}~\bibnamefont{Bloch}},
		\bibinfo{journal}{Science} \textbf{\bibinfo{volume}{357}},
		\bibinfo{pages}{995} (\bibinfo{year}{2017}),
		\urlprefix\url{https://doi.org/10.1126/science.aal3837}.
		
		\bibitem[{\citenamefont{Sherson et~al.}(2010)\citenamefont{Sherson, Weitenberg,
				Endres, Cheneau, Bloch, and Kuhr}}]{sherson}
		\bibinfo{author}{\bibfnamefont{J.~F.} \bibnamefont{Sherson}},
		\bibinfo{author}{\bibfnamefont{C.}~\bibnamefont{Weitenberg}},
		\bibinfo{author}{\bibfnamefont{M.}~\bibnamefont{Endres}},
		\bibinfo{author}{\bibfnamefont{M.}~\bibnamefont{Cheneau}},
		\bibinfo{author}{\bibfnamefont{I.}~\bibnamefont{Bloch}}, \bibnamefont{and}
		\bibinfo{author}{\bibfnamefont{S.}~\bibnamefont{Kuhr}},
		\bibinfo{journal}{Nature} \textbf{\bibinfo{volume}{467}}, \bibinfo{pages}{68}
		(\bibinfo{year}{2010}), \urlprefix\url{https://doi.org/10.1038/nature09378}.
		
		\bibitem[{\citenamefont{Fukuhara et~al.}(2015)\citenamefont{Fukuhara, Hild,
				Zeiher, Schau\ss{}, Bloch, Endres, and Gross}}]{fukuhara}
		\bibinfo{author}{\bibfnamefont{T.}~\bibnamefont{Fukuhara}},
		\bibinfo{author}{\bibfnamefont{S.}~\bibnamefont{Hild}},
		\bibinfo{author}{\bibfnamefont{J.}~\bibnamefont{Zeiher}},
		\bibinfo{author}{\bibfnamefont{P.}~\bibnamefont{Schau\ss{}}},
		\bibinfo{author}{\bibfnamefont{I.}~\bibnamefont{Bloch}},
		\bibinfo{author}{\bibfnamefont{M.}~\bibnamefont{Endres}}, \bibnamefont{and}
		\bibinfo{author}{\bibfnamefont{C.}~\bibnamefont{Gross}},
		\bibinfo{journal}{Phys. Rev. Lett.} \textbf{\bibinfo{volume}{115}},
		\bibinfo{pages}{035302} (\bibinfo{year}{2015}),
		\urlprefix\url{https://link.aps.org/doi/10.1103/PhysRevLett.115.035302}.
		
		\bibitem[{\citenamefont{Fukuhara et~al.}(2013)\citenamefont{Fukuhara, Kantian,
				Endres, Cheneau, Schau{\ss}, Hild, Bellem, Schollw{\"o}ck, Giamarchi, Gross
				et~al.}}]{fukuhara2}
		\bibinfo{author}{\bibfnamefont{T.}~\bibnamefont{Fukuhara}},
		\bibinfo{author}{\bibfnamefont{A.}~\bibnamefont{Kantian}},
		\bibinfo{author}{\bibfnamefont{M.}~\bibnamefont{Endres}},
		\bibinfo{author}{\bibfnamefont{M.}~\bibnamefont{Cheneau}},
		\bibinfo{author}{\bibfnamefont{P.}~\bibnamefont{Schau{\ss}}},
		\bibinfo{author}{\bibfnamefont{S.}~\bibnamefont{Hild}},
		\bibinfo{author}{\bibfnamefont{D.}~\bibnamefont{Bellem}},
		\bibinfo{author}{\bibfnamefont{U.}~\bibnamefont{Schollw{\"o}ck}},
		\bibinfo{author}{\bibfnamefont{T.}~\bibnamefont{Giamarchi}},
		\bibinfo{author}{\bibfnamefont{C.}~\bibnamefont{Gross}},
		\bibnamefont{et~al.}, \bibinfo{journal}{Nature Physics}
		\textbf{\bibinfo{volume}{9}}, \bibinfo{pages}{235} (\bibinfo{year}{2013}),
		\urlprefix\url{https://doi.org/10.1038/nphys2561}.
		
		\bibitem[{\citenamefont{Subrahmanyam}(2004)}]{spindynamics1}
		\bibinfo{author}{\bibfnamefont{V.}~\bibnamefont{Subrahmanyam}},
		\bibinfo{journal}{Phys. Rev. A} \textbf{\bibinfo{volume}{69}},
		\bibinfo{pages}{034304} (\bibinfo{year}{2004}),
		\urlprefix\url{https://link.aps.org/doi/10.1103/PhysRevA.69.034304}.
		
		\bibitem[{\citenamefont{Amico et~al.}(2004)\citenamefont{Amico, Osterloh,
				Plastina, Fazio, and Massimo~Palma}}]{spindynamics2}
		\bibinfo{author}{\bibfnamefont{L.}~\bibnamefont{Amico}},
		\bibinfo{author}{\bibfnamefont{A.}~\bibnamefont{Osterloh}},
		\bibinfo{author}{\bibfnamefont{F.}~\bibnamefont{Plastina}},
		\bibinfo{author}{\bibfnamefont{R.}~\bibnamefont{Fazio}}, \bibnamefont{and}
		\bibinfo{author}{\bibfnamefont{G.}~\bibnamefont{Massimo~Palma}},
		\bibinfo{journal}{Phys. Rev. A} \textbf{\bibinfo{volume}{69}},
		\bibinfo{pages}{022304} (\bibinfo{year}{2004}),
		\urlprefix\url{https://link.aps.org/doi/10.1103/PhysRevA.69.022304}.
		
		\bibitem[{\citenamefont{Duarte et~al.}(2017)\citenamefont{Duarte, Carvalho,
				Bernardes, and de~Melo}}]{duarte}
		\bibinfo{author}{\bibfnamefont{C.}~\bibnamefont{Duarte}},
		\bibinfo{author}{\bibfnamefont{G.~D.} \bibnamefont{Carvalho}},
		\bibinfo{author}{\bibfnamefont{N.~K.} \bibnamefont{Bernardes}},
		\bibnamefont{and} \bibinfo{author}{\bibfnamefont{F.}~\bibnamefont{de~Melo}},
		\bibinfo{journal}{Phys. Rev. A} \textbf{\bibinfo{volume}{96}},
		\bibinfo{pages}{032113} (\bibinfo{year}{2017}),
		\urlprefix\url{https://link.aps.org/doi/10.1103/PhysRevA.96.032113}.
		
		\bibitem[{\citenamefont{Correia}(2017)}]{PedroM2017}
		\bibinfo{author}{\bibfnamefont{P.~S.} \bibnamefont{Correia}}, Master's thesis,
		\bibinfo{school}{Centro Brasileiro de Pesquisas Físicas},
		\bibinfo{address}{Rio de Janeiro - RJ} (\bibinfo{year}{2017}),
		\urlprefix\url{http://cbpfindex.cbpf.br/publication_pdfs/dissertacaoDeMestrado_2018-11-06-16-43-19ZGlzc2VydGFjYW9EZU1lc3RyYWRv.pdf}.
		
		\bibitem[{\citenamefont{Ott}(2016)}]{atomdet}
		\bibinfo{author}{\bibfnamefont{H.}~\bibnamefont{Ott}},
		\bibinfo{journal}{Reports on Progress in Physics}
		\textbf{\bibinfo{volume}{79}}, \bibinfo{pages}{054401}
		(\bibinfo{year}{2016}),
		\urlprefix\url{https://doi.org/10.1088%2F0034-4885%2F79%2F5%2F054401}.
			
			\bibitem[{\citenamefont{Kadanoff}(1966)}]{kadanoff}
			\bibinfo{author}{\bibfnamefont{L.~P.} \bibnamefont{Kadanoff}},
			\bibinfo{journal}{Physics Physique Fizika} \textbf{\bibinfo{volume}{2}},
			\bibinfo{pages}{263} (\bibinfo{year}{1966}),
			\urlprefix\url{https://link.aps.org/doi/10.1103/PhysicsPhysiqueFizika.2.263}.
			
			\bibitem[{\citenamefont{Wilson}(1975)}]{wilson}
			\bibinfo{author}{\bibfnamefont{K.~G.} \bibnamefont{Wilson}},
			\bibinfo{journal}{Rev. Mod. Phys.} \textbf{\bibinfo{volume}{47}},
			\bibinfo{pages}{773} (\bibinfo{year}{1975}),
			\urlprefix\url{https://link.aps.org/doi/10.1103/RevModPhys.47.773}.
			
			\bibitem[{\citenamefont{Evenbly and Vidal}(2014)}]{vidal2007entanglement}
			\bibinfo{author}{\bibfnamefont{G.}~\bibnamefont{Evenbly}} \bibnamefont{and}
			\bibinfo{author}{\bibfnamefont{G.}~\bibnamefont{Vidal}},
			\bibinfo{journal}{Phys. Rev. Lett.} \textbf{\bibinfo{volume}{112}},
			\bibinfo{pages}{220502} (\bibinfo{year}{2014}),
			\urlprefix\url{https://link.aps.org/doi/10.1103/PhysRevLett.112.220502}.
			
			\bibitem[{\citenamefont{Saideh et~al.}(2015)\citenamefont{Saideh, Ribeiro,
					Ferrini, Coudreau, Milman, and Keller}}]{Ibrahim}
			\bibinfo{author}{\bibfnamefont{I.}~\bibnamefont{Saideh}},
			\bibinfo{author}{\bibfnamefont{A.~D.} \bibnamefont{Ribeiro}},
			\bibinfo{author}{\bibfnamefont{G.}~\bibnamefont{Ferrini}},
			\bibinfo{author}{\bibfnamefont{T.}~\bibnamefont{Coudreau}},
			\bibinfo{author}{\bibfnamefont{P.}~\bibnamefont{Milman}}, \bibnamefont{and}
			\bibinfo{author}{\bibfnamefont{A.}~\bibnamefont{Keller}},
			\bibinfo{journal}{Phys. Rev. A} \textbf{\bibinfo{volume}{92}},
			\bibinfo{pages}{052334} (\bibinfo{year}{2015}),
			\urlprefix\url{https://link.aps.org/doi/10.1103/PhysRevA.92.052334}.
			
			\bibitem[{\citenamefont{Bloch}(2005)}]{bloch}
			\bibinfo{author}{\bibfnamefont{I.}~\bibnamefont{Bloch}},
			\bibinfo{journal}{Nature Physics} \textbf{\bibinfo{volume}{1}},
			\bibinfo{pages}{23} (\bibinfo{year}{2005}),
			\urlprefix\url{https://doi.org/10.1038/nphys138}.
			
			\bibitem[{\citenamefont{Bloch}(2004)}]{bloch2}
			\bibinfo{author}{\bibfnamefont{I.}~\bibnamefont{Bloch}},
			\bibinfo{journal}{Physics World} \textbf{\bibinfo{volume}{17}},
			\bibinfo{pages}{25} (\bibinfo{year}{2004}),
			\urlprefix\url{https://physicsworld.com/a/quantum-gases-in-optical-lattices/}.
			
			\bibitem[{\citenamefont{Duan et~al.}(2003)\citenamefont{Duan, Demler, and
					Lukin}}]{heisenbergxx}
			\bibinfo{author}{\bibfnamefont{L.-M.} \bibnamefont{Duan}},
			\bibinfo{author}{\bibfnamefont{E.}~\bibnamefont{Demler}}, \bibnamefont{and}
			\bibinfo{author}{\bibfnamefont{M.~D.} \bibnamefont{Lukin}},
			\bibinfo{journal}{Phys. Rev. Lett.} \textbf{\bibinfo{volume}{91}},
			\bibinfo{pages}{090402} (\bibinfo{year}{2003}),
			\urlprefix\url{https://link.aps.org/doi/10.1103/PhysRevLett.91.090402}.
			
			\bibitem[{\citenamefont{Konno}(2005)}]{bessel}
			\bibinfo{author}{\bibfnamefont{N.}~\bibnamefont{Konno}},
			\bibinfo{journal}{Phys. Rev. E} \textbf{\bibinfo{volume}{72}},
			\bibinfo{pages}{026113} (\bibinfo{year}{2005}),
			\urlprefix\url{https://link.aps.org/doi/10.1103/PhysRevE.72.026113}.
			
			\bibitem[{\citenamefont{Yu and Eberly}(2007)}]{concX}
			\bibinfo{author}{\bibfnamefont{T.}~\bibnamefont{Yu}} \bibnamefont{and}
			\bibinfo{author}{\bibfnamefont{J.}~\bibnamefont{Eberly}},
			\bibinfo{journal}{Quant. Info. Comp.} \textbf{\bibinfo{volume}{7}},
			\bibinfo{pages}{459} (\bibinfo{year}{2007}),
			\urlprefix\url{https://doi.org/10.26421/QIC7.5-6}.
			
			\bibitem[{\citenamefont{Mazza et~al.}(2015)\citenamefont{Mazza, Rossini, Fazio,
					and Endres}}]{mazza}
			\bibinfo{author}{\bibfnamefont{L.}~\bibnamefont{Mazza}},
			\bibinfo{author}{\bibfnamefont{D.}~\bibnamefont{Rossini}},
			\bibinfo{author}{\bibfnamefont{R.}~\bibnamefont{Fazio}}, \bibnamefont{and}
			\bibinfo{author}{\bibfnamefont{M.}~\bibnamefont{Endres}},
			\bibinfo{journal}{New Journal of Physics} \textbf{\bibinfo{volume}{17}},
			\bibinfo{pages}{013015} (\bibinfo{year}{2015}),
			\urlprefix\url{https://doi.org/10.1088/1367-2630/17/1/013015}.
			
			\bibitem[{\citenamefont{Vidal and Werner}(2002)}]{Neg2002}
			\bibinfo{author}{\bibfnamefont{G.}~\bibnamefont{Vidal}} \bibnamefont{and}
			\bibinfo{author}{\bibfnamefont{R.~F.} \bibnamefont{Werner}},
			\bibinfo{journal}{Phys. Rev. A} \textbf{\bibinfo{volume}{65}},
			\bibinfo{pages}{032314} (\bibinfo{year}{2002}),
			\urlprefix\url{https://link.aps.org/doi/10.1103/PhysRevA.65.032314}.
			
			\bibitem[{\citenamefont{Iemini et~al.}(2016)\citenamefont{Iemini, Russomanno,
					Rossini, Scardicchio, and Fazio}}]{referee1}
			\bibinfo{author}{\bibfnamefont{F.}~\bibnamefont{Iemini}},
			\bibinfo{author}{\bibfnamefont{A.}~\bibnamefont{Russomanno}},
			\bibinfo{author}{\bibfnamefont{D.}~\bibnamefont{Rossini}},
			\bibinfo{author}{\bibfnamefont{A.}~\bibnamefont{Scardicchio}},
			\bibnamefont{and} \bibinfo{author}{\bibfnamefont{R.}~\bibnamefont{Fazio}},
			\bibinfo{journal}{Phys. Rev. B} \textbf{\bibinfo{volume}{94}},
			\bibinfo{pages}{214206} (\bibinfo{year}{2016}),
			\urlprefix\url{https://link.aps.org/doi/10.1103/PhysRevB.94.214206}.
			
			\bibitem[{\citenamefont{Wang et~al.}(2018)\citenamefont{Wang, Liu, and
					Hu}}]{referee2}
			\bibinfo{author}{\bibfnamefont{J.}~\bibnamefont{Wang}},
			\bibinfo{author}{\bibfnamefont{X.-J.} \bibnamefont{Liu}}, \bibnamefont{and}
			\bibinfo{author}{\bibfnamefont{H.}~\bibnamefont{Hu}}, \bibinfo{journal}{New
				Journal of Physics} \textbf{\bibinfo{volume}{20}}, \bibinfo{pages}{053015}
			(\bibinfo{year}{2018}),
			\urlprefix\url{https://doi.org/10.1088%2F1367-2630%2Faabe3d}.
				
\end{thebibliography}

\end{document}